\documentclass[%
 prb,
 sd,%
 amsmath,amssymb,
 reprint,%
]{revtex4-1}

\usepackage{graphicx}% Include figure files
\usepackage{dcolumn}% Align table columns on decimal point
\usepackage{bm}% bold math
\usepackage{booktabs} %For Tables
\begin{document}

\preprint{AIP/123-QED}
\title[Spin-triplet supercurrent in Josephson junctions containing a synthetic antiferromagnet with perpendicular magnetic anisotropy]{Spin-triplet supercurrent in Josephson junctions containing a synthetic antiferromagnet with perpendicular magnetic anisotropy
}

\author{Joseph A. Glick}
\author{Samuel Edwards}
\author{Demet Korucu}\altaffiliation[Permanent address: ]{Department of Physics, Gazi University, Ankara, Turkey}
\author{Victor Aguilar}
\author{Bethany M. Niedzielski}
\author{Reza Loloee}
\author{W. P. Pratt, Jr.}
\author{Norman O. Birge}\email{birge@pa.msu.edu}
\affiliation{Department of Physics and Astronomy, Michigan State University, East Lansing, MI 48824, USA}
\author{P. G. Kotula}
\author{N. Missert}
\affiliation{Sandia National Laboratories, Albuquerque, NM 87185, USA}

\date{\today}

\begin{abstract}
 We present measurements of Josephson junctions containing three magnetic layers with noncolinear magnetizations. The junctions are of the form $S/F^{\prime}/N/F/N/F^{\prime \prime}/S$, where $S$ is superconducting Nb, $F^\prime$ is either a thin Ni or Permalloy layer with in-plane magnetization, $N$ is the normal metal Cu, $F$ is a synthetic antiferromagnet (SAF) with magnetization perpendicular to the plane, composed of Pd/Co multilayers on either side of a thin Ru spacer, and $F^{\prime \prime}$ is a thin Ni layer with in-plane magnetization. The supercurrent in these junctions decays more slowly as a function of the $F$-layer thickness than for similar spin-singlet junctions not containing the $F^\prime$ and $F^{\prime \prime}$ layers. The slower decay is the prime signature that the supercurrent in the central part of these junctions is carried by spin-triplet pairs.  The junctions containing $F^{\prime}=$ Permalloy are suitable for future experiments where either the amplitude of the critical current or the ground-state phase difference across the junction is controlled by changing the relative orientations of the magnetizations of the $F^{\prime}$ and $F^{\prime \prime}$ layers.

%Valid PACS numbers may be entered using the \verb+\pacs{#1}+ command.
\end{abstract}

\pacs{ }% PACS, the Physics and Astronomy Classification Scheme.
%Use showkeys class option if keyword
\maketitle

\section{\label{sec:level1}Introduction}
When a conventional superconductor is brought into contact with a ferromagnetic material, Cooper pairs will penetrate into the ferromagnet via the proximity effect and remain correlated over length scales typically only a few nanometers~\cite{Buzdin_SFReview2005}. The exchange field in the ferromagnet causes spin-up and spin-down electrons to split into different bands that have different momenta at the Fermi level. Thus opposite spin electron pairs acquire a net center-of-mass momentum-- i.e. the pair-correlation function does not only decay, but oscillates in sign as a function of the ferromagnetic layer thickness ~\cite{Fulde1964, Larkin1964, Buzdin1982, Demler1997}. Experimental verification of this fact was achieved in the early 2000's~\cite{Ryazanov2001, Kontos2002}, followed by many other groups who measured the oscillation and decay of the critial current in Josephson junctions~\cite{Blum2002,Sellier2003,Shelukhin2006,Weides2006,Robinson2006,Robinson2007,Bannykh2009,Khaire2009,Niedzielski2015,Glick2016}. In these systems the resulting pair correlation function in the ferromagnet contains a mixture of both spin-singlet and spin-triplet components with magnetic quantum number $m$ = 0~\cite{Eschrig2003}.

A thoroughly different situation occurs if the electrons near the superconducting/ferromagnetic (S/F) interface are paired with the same spin-orientation. In that case, both electrons enter the same spin band and can remain correlated in the ferromagnet over much longer distances. In a Josephson junction there is thus an overall slower decay in the critical current and no oscillation of the supercurrent with the thickness of the ferromagnet~\cite{Bergeret2001}. While spin-triplet superconductivity is not commonly found in nature, it was predicted~\cite{Bergeret2001, Kadigrobov2001, Eschrig2003, Volkov2003, Bergeret2005} that it could be engineered in multi-layered ferromagnetic systems having noncolinear magnetizations or magnetic inhomogeneity.  Since then, demonstrations of spin-triplet proximity effects have been reported by many groups using a variety of experimental techniques~\cite{Keizer2006, Sosnin2006, Khaire2010, Robinson2010Science, Sprungmann2010, Anwar2010, Wang2010, Klose2012, Leksin2012, Zdravkov2013, Banerjee2014-TripletSpinValve, Banerjee2014, Iovan2014,  Wang2014, Jara2014, Flokstra2015, Singh2015, Linder2015, Eschrig2015, DiBernardo2015, Feng2017, Cirillo2017}.

Our group has focussed on spin-triplet Josephson junctions (JJs) of the form suggested by Houzet \& Buzdin \cite{Houzet2007}, $S/F^{\prime}/N/F/N/F^{\prime \prime}/S$~\cite{Khaire2010, Klose2012, Gingrich2012, Martinez2016}. Such a JJ converts between spin-singlet and long-range spin-triplet supercurrent in the following manner~\cite{Eschrig2011}: 1) spin-singlet pairs from the first superconductor enter the first ferromagnet ($F^\prime$) and acquire a net center-of-mass momentum, generating a short-range ($m = 0$) spin-triplet component, as previously described.  2) The electron pairs enter a second ferromagnet ($F$) with magnetization non-colinear to the first. Expression of the $m = 0$ triplet state in the rotated basis includes the long-range ($m = \pm 1$) spin-triplet states; the spin-triplet conversion is maximized when the magnetizations are perpendicular. While the spin-singlet and $m = 0$ spin-triplet states decay quickly in $F$, the ($m = \pm 1$) states propagate much further. 3) Since the final superconducting electrode can only accept spin-singlet states, a third ferromagnetic layer ($F^{\prime \prime}$) is needed to convert the long-range triplet states back into spin-singlet. Again, that conversion process is optimal when the magnetization of $F^{\prime \prime}$ is perpendicular to that of $F$.  Recently, our group successfully implemented a scheme where by rotating the magnetization of the $F^{\prime \prime}$ layer, the long-range triplet supercurrent could be controllably toggled ``on'' or ``off'' as evidenced by large amplitude changes in the critical current~\cite{Martinez2016}.

The theory of spin-triplet JJs predicts that the ground-state phase difference across a junction of the form described above depends on the relative orientations of the three magnetizations~\cite{Volkov2003, Eschrig2003, Bergeret2005, Houzet2007, Volkov2010, Trifunovic2010}. Spin-triplet junctions where the magnetizations of all three ferromagnetic layers are coplanar exhibit complementary 0 and $\pi$-phase states dependent only on whether the outer two magnetizations are parallel or antiparallel.

In principle, there are many ways to design a JJ to test that prediction. We have focused on designs where the magnetizations of the outer $F^\prime$ and $F^{\prime \prime}$ layers both lie in-plane; the junctions are given an elliptical shape to set the directions of those in-plane magnetizations by shape anisotropy. But the elliptical junction shape makes it difficult to achieve non-colinear magnetization in the central $F$ layer, if it is also in-plane. A solution is to use out-of-plane magnetization for $F$, which is easily accomplished using a magnetic material with strong perpendicular magnetic anisotropy (PMA)~\cite{Gingrich2012}. Then, one can utilize shape anisotropy to preferentially orient the magnetization direction of the $F^\prime$ and $F^{\prime \prime}$ layers, all the while preserving the optimal 90 degree relative magnetization angle between each successive ferromagnetic layer. Previous efforts by our group to detect the JJ phase change using such a design were only partially successful, however~\cite{GingrichThesis2014}; while a $\pi$ phase change appeared in some experiments, the magnetic behavior of the junctions was poor and irreproducible. We suspected at the time that stray fields from the domain walls in the PMA $F$ layer penetrated the $F^\prime$ and $F^{\prime \prime}$ layers and ruined their magnetic properties. A possible solution to that problem is to replace the central PMA layer with a PMA synthetic antiferromagnet (SAF), in which each magnetic domain in the lower half of the SAF is coupled to a domain with opposite-pointing magnetization in the upper half of the SAF. Such a system should produce minimal stray fields in the $F^\prime$ and $F^{\prime \prime}$ layers that are located above and below the SAF~\cite{SchneiderPrivatecomm}.

Here we report measurements of $S/F^{\prime}/N/F/N/F^{\prime \prime}/S$ spin-triplet based Josephson junctions in which the central magnetic layer, $F$, is a synthetic antiferromagnet (SAF) based on [Pd/Co] multilayers with PMA. This study was performed with an eye towards future devices in which the phase state of the junctions can be reliably controlled. The main result of this work is that the critical current in these junctions decays more slowly with increasing thickness of the PMA SAF than it does in junctions that do not contain the $F^\prime$ and $F^{\prime \prime}$ layers. That result represents strong evidence that the supercurrent in the central part of these junctions is carried by spin-triplet pairs, whereas it is carried only by the short-range components in the control samples.  These junctions are therefore suitable for phase-control experiments envisioned above.

\section{\label{sec:level2}Sample Fabrication and Characterization}
\subsection{Magnetic Properties of Pd/Co Multilayers and Synthetic Antiferromagnets With Perpendicular Magnetic Anisotropy}
Thin multilayers of Pd and Co can be grown to have perpendicular magnetic anisotropy (PMA), i.e. with magnetization perpendicular to the sample plane, as reported by Chang \textit{et al.}~\cite{Chang2013}. Moreover, when two such Pd/Co multilayers are separated by a thin normal metal spacer (such as Ru, Rh, Ir, or Cu), they may couple via the exchange interaction to form a SAF in which the magnetizations on either side of the spacer align in an antiparallel fashion. The outstanding PMA and SAF properties of such layers have attracted interest towards their use in spin-transfer-torque magnetic random access memories~\cite{Lee2016} and other applications. We investigate if PMA SAFs could be advantageous for spin-triplet JJs with phase control. For the PMA SAF to serve as the central $F$ layer and optimize the generation of long-range spin-triplet supercurrent its magnetization needs to remain pinned perpendicular to the sample plane over the range of measurement fields used in the experiments.

To characterize the Pd/Co multilayers and verify that they have PMA, we sputtered films of: Nb(5)/Cu(5)/[Pd(d$_{\mathrm{Pd}}$)/Co(0.3)]$_n$/Pd(d$_{\mathrm{Pd}}$)/Cu(5)/Nb(5), where the layer thicknesses in nanometers are shown in parentheses and the sequences in brackets are repeated $n$ = 10 or 20 times. Similar to Chang \textit{et al.}~\cite{Chang2013} we fix the Co thickness to 0.3 nm while the Pd thickness, d$_{\mathrm{Pd}}$, was varied from 0.8 - 1.0 nm.

All the samples throughout this paper were fabricated using high-vacuum sputtering deposition on 0.5$\times$0.5 in$^2$ silicon chips as follows. Before sputtering, the chamber was baked for eight hours and reduced to a base pressure of 2$\times$10$^{-8}$ Torr with a cryopump. The chamber was then cooled by circulating liquid nitrogen though a Meissner trap to reduce the partial pressure of water in the chamber to $<$ 3 $\times$ 10$^{-9}$ Torr as confirmed by an \textit{in-situ} residual gas analyzer. The films were deposited via dc sputtering with either 1-inch magnetron or 2.25-inch triode guns in an Argon plasma of pressure 1.3 $\times$ 10$^{-3}$ Torr. During the deposition the sample temperature was held between $-30\,^{\circ}\mathrm{C}$ and $-20\,^{\circ}\mathrm{C}$. The thicknesses of the various deposited materials were controlled by measuring the deposition rates (accurate to $\pm 0.1 \mathrm{\AA}$/s) using a crystal film thickness monitor and a computer controlled stepper motor that operates the position of the shutter and sample plate.

\begin{figure}
	\begin{center}
		\includegraphics[width=\linewidth]{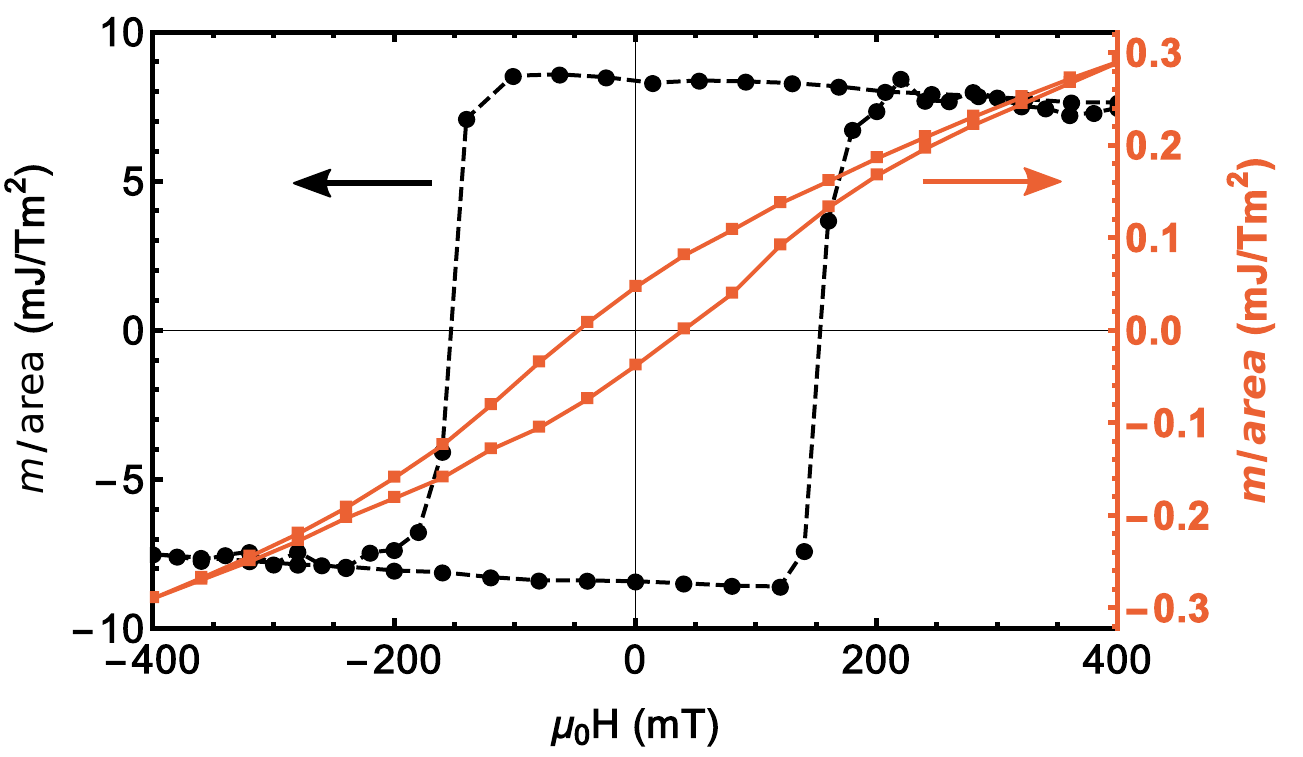}
	\end{center}
	\caption{\label{fig:M-Hloop_PdCo_10Repeats} Hysteresis loops of a Pd/Co multilayer film measured at 5K using a SQUID magnetometer. The data are expressed as total magnetic moment $m$ per unit area. With the applied magnetic field perpendicular to the sample plane (black, left axis), the loop is square-like indicating the Pd/Co multilayer has perpendicular magnetic anisotropy (PMA), with a large coercive field of over 160 mT. With the applied field parallel to the sample plane (red, right axis) the hysteresis is only slightly discernable, thus the magnetization has a very small in-plane component. Also, note the comparatively smaller scale on the right axis. The sample is composed of Nb(5)/Cu(5)/[Pd(0.9)/Co(0.3)]$_{10}$/Pd(0.9)/Cu(5)/Nb(5) with thicknesses in nanometers. Lines are to guide the eye.}
\end{figure}

We measured the films' magnetic moment $m$ per unit area vs. the applied magnetic field ($M$-$H$ loop), using a Quantum Design dc-SQUID magnetometer at 5 K. Sweeping an applied magnetic field that is perpendicular to the sample plane results in square-like  $M$-$H$ loops, as shown in Fig.~\ref{fig:M-Hloop_PdCo_10Repeats} (black data points), confirming that the Pd/Co multilayer has PMA. The sample with d$_\mathrm{Pd}$ = 0.9 nm had the best magnetic properties: a coercive field of over 160 mT and the largest squareness. With $H$ applied parallel to the sample plane, the $M$-$H$ loops show only a slight hysteresis with small remanent magnetization, indicating a very small in-plane moment (red data points). Note that dividing the $m/\mathrm{area}$ values by the total Co thickness of 3.0 nm gives a saturation magnetization of about 2.7$\times$10$^6$ A/m (=2700 emu/cm$^3$) which is about twice the saturation magnetization of bulk Co. This is because the Co partially polarizes the surrounding Pd layers~\cite{Smith2008}.

Next, we characterized the magnetic behavior of two such Pd/Co multilayers arranged on either side of a thin Ru spacer to form a SAF with PMA. The coupling of the Pd/Co multilayer into a SAF structure arises due to interlayer exchange coupling (IEC) between the two multilayers. The energy density of the IEC can be modulated by tuning the thickness of the spacer layer and depends strongly upon which material(s) it forms interfaces with. We arranged the Ru spacer to have adjoining Co layers on either side, similar to Chang \textit{et al.}~\cite{Chang2013}.

\begin{figure}
	\begin{center}
		\includegraphics[width=0.93\linewidth]{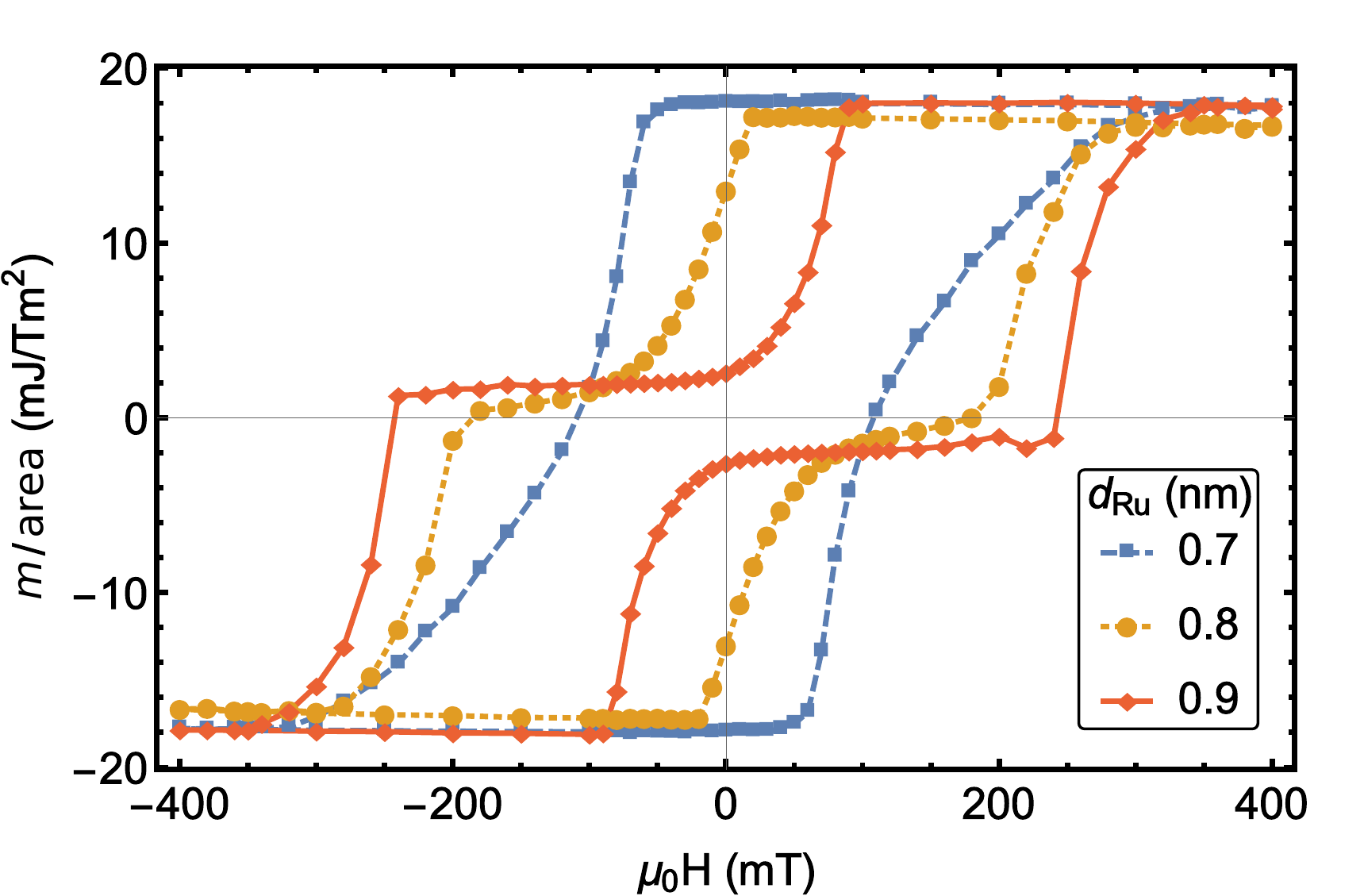}
	\end{center}
	\caption{\label{fig:M-Hloop_PdCo_SAF} Hysteresis loop measurements of synthetic antiferromagnet (SAFs) films measured using a dc-SQUID magnetometer at 5K with the applied field perpendicular to the sample plane. The data are expressed as total magnetic moment $m$ per unit area. The SAF is composed of Pd(0.9 nm)/Co(0.3 nm) multilayers separated by a thin Ru spacer, whose thickness, $d_{\mathrm{Ru}}$, was varied from 0.7 to 1.1 nm (0.7-0.9 nm shown). At a Ru thickness of 0.7 nm (blue) the Pd/Co multilayers are ferromagnetically coupled out-of-plane, since only a single loop is observable. However, as the Ru thickness increases, the out-of-plane $M$-$H$ loops have an intermediate step with a flat plateau in the magnetization (yellow, red).  The width of the intermediate plateau is maximal near d$_{\mathrm{Ru}}$ = 0.9 (red), indicating stable antiferromagnetic exchange coupling at applied fields less than $\pm$ 250 mT. Lines are to guide the eye.
}
\end{figure}

To optimize the antiferromagnetic coupling in the SAF, we sputtered a set of samples with an ``unbalanced'' SAF configuration of the form: Nb(5)/Cu(5)/[Pd(0.9)/Co(0.3)]$_{12}$/Ru($d_{\mathrm{Ru}}$)/ [Pd(0.9)/Co(0.3)]$_{10}$/Cu(5)/Nb(5), varying the Ru thickness between $d_{\mathrm{Ru}}$=0.7, 0.8, ..., 1.1 nm. We measured the samples' magnetic response in a dc-SQUID magnetometer with $H$ perpendicular to the sample plane. As shown in Fig.~\ref{fig:M-Hloop_PdCo_SAF}, at a Ru thickness of 0.7 nm the Pd/Co multilayers are ferromagnetically coupled, since only a single (slightly distorted) loop is observable. As the Ru thickness increases, the $M$-$H$ loops have an intermediate step with a flat plateau in the magnetization, indicating stable antiferromagnetic exchange coupling at applied fields less than $\pm$ 250 mT. The maximum width of the intermediate plateau, and hence the maximum antiferromagnetic coupling measured, was obtained for the samples with d$_{\mathrm{Ru}}$ = 0.9 and 1.0 nm (not shown). Therefore, in the Josephson junction samples presented in the next section we choose to fix d$_{\mathrm{Ru}}$ = 0.95 nm.

Note that the extra two repeats of the Pd/Co multilayer in the data presented in Fig.~\ref{fig:M-Hloop_PdCo_SAF} are added merely to accentuate the separation between the two corresponding hysteresis loops, allowing us to more easily determine the optimal Ru thickness. In spin-triplet JJs, it is desirable to maximize the flux cancelation within the PMA SAF. Thus, in the experiments that follow, we used a balanced SAF structure which has an equal total thickness of Pd and Co on either side of the Ru spacer.

We briefly mention that we tested another, similar type of PMA SAF which was composed of Ni/Co multilayers of the form: [Co(0.3)/Ni(0.6)]$_{n}$/Co(0.3)/ Ru($d_{\mathrm{Ru}}$)/[Co(0.3)/Ni(0.6)]$_{m}$ /Co(0.3), where $n$ = 4 and $m$ = 3 or 4. The magnetic behavoir of these PMA SAFs was quite similar to the data in Fig.~\ref{fig:M-Hloop_PdCo_SAF}, with strong antiferromagnetic pinning, PMA, and square-shaped $M$-$H$ loops. However, after fabricating them into Josephson junctions, we found the Ni/Co SAFs to be rather unsatisfactory from a device perspective in that they suffered from extremely small critical current. The same was not true for the Pd/Co based PMA SAFs which will be discussed in the next section.
\subsection{Josephson Junctions}
\begin{figure}
	\begin{center}
		\includegraphics[width=\linewidth]{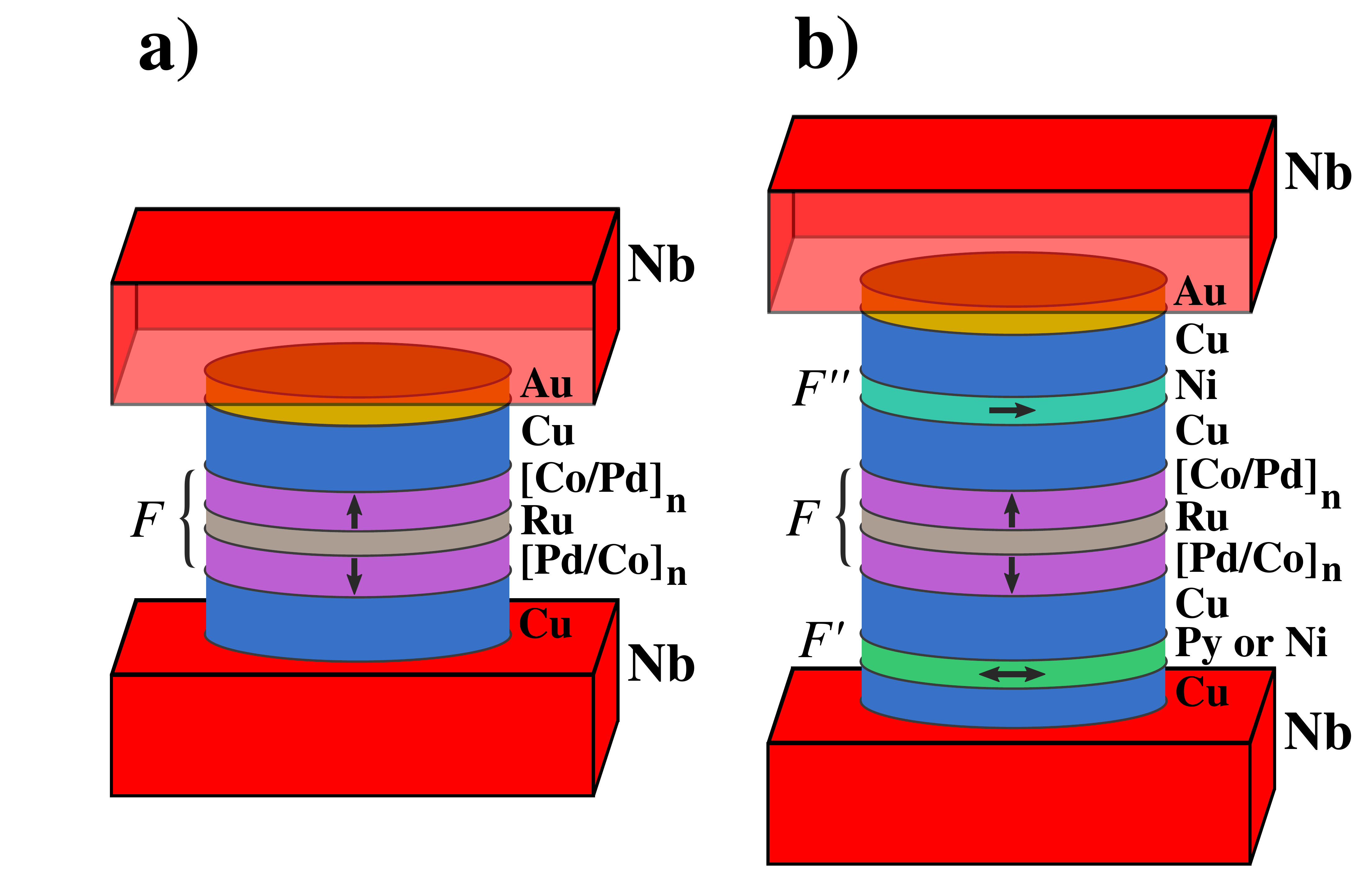}
	\end{center}
	\caption{\label{fig:SampleStructure} A schematic representation of the vertical cross sectional structure of our Josephson junctions (not to scale). The central $F$ layer is composed of two sets of $n$ [Pd(0.9 mm)/Co(0.3) nm] bilayers with perpendicular magnetic anisotropy (PMA), on either side of a Ru(0.95 nm) spacer to form a synthetic antiferromagnet (SAF).
	(a) With only the PMA SAF in the center, the supercurrent is carried by short-range spin-singlet pairs. (b) When combined with the two other ferromagnets, F$^\prime$ and F$^{\prime \prime}$, with in-plane magnetization, the supercurrent in F is carried by long-range spin-triplet pairs.  In this study, the $F^\prime$ layer is either Ni(1.6 nm), which maximizes the spin-triplet supercurrent, or Py (1.25 nm), which can act as a ``free'' layer, switching its magnetization at a low field.}
\end{figure}

We next seek to address the following questions: 1) how does the Pd/Co PMA SAF structure behave as a barrier to current transport in Josephson junctions (JJs)? 2) can a PMA SAF of this nature be utilized in a JJ device with ferromagnetic layers to generate spin-triplet supercurrent? Both questions can be answered by measuring how the critial current in these type of junctions varies with the number of Pd/Co layers on either side of the Ru spacer. To this end we fabricated three sets of JJs. The first is a series of control samples which contain only the Pd/Co PMA SAF, shown in Fig.~\ref{fig:SampleStructure} (a), and are meant to measure the decay of the short-range spin-singlet supercurrent. The second and third series of JJs are designed to carry long-range spin-triplet supercurrent. They contain a Pd/Co PMA SAF centered between two additional ferromagnetic layers with in-plane magnetization, shown in Fig.~\ref{fig:SampleStructure} (b). In the second set both the bottom ferromagnet, $F^\prime$, and the top ferromagent, $F^{\prime \prime}$, are Ni with thickness 1.6 nm. The third set of JJ's is similar, but the $F^\prime$ layer is Permalloy (Py = Ni$_{81}$Fe$_{19}$) with thickness 1.25 nm and the $F^{\prime \prime}$ layer is Ni(1.6 nm), as shown Fig.~\ref{fig:SampleStructure} (b).

Due to Permalloy's sharp magnetic switching at low magnetic field, we envision using it in controllable JJs~\cite{Baek2014, Gingrich2016, Martinez2016, Niedzielski2017}. However, from previous experience we know that Py's magnetic properties degrade if grown on a rough surface. Thus, in the third series of JJ's, the Py layer is intentionally placed near the bottom of the stack where it will be the least effected by upward-propagating surface roughness, as discussed in more detail later in Fig.~\ref{fig:SandiaSTEM}. Nickel, while harder to control magnetically, is the ferromagnetic material we and others have found to be the least detrimental to the propagation of supercurrent~\cite{Baek2017}. Thus, it was used as both $F^\prime$ and $F^{\prime \prime}$ in the second set of samples in case the critical currents in the JJs with Py were too small to measure. The role of the other layers in Fig.~\ref{fig:SampleStructure} will be discussed later.

\subsection{Josephson Junction Sample Fabrication}
The sample fabrication proceeds similarly to other nanopillar junctions made by our group~\cite{Niedzielski2015, Glick2016}, which we outline in somewhat less detail here, but specifically mention any differences. The geometry of the bottom leads was defined via optical photolithography and a lift-off process. The bottom electrode is a sputtered Nb/Al multilayer of form [Nb(25)/Al(2.4)]$_3$/Nb(20), which is much smoother than a continuous Nb layer~\cite{Wang2012, Thomas1998, Kohlstedt1996}, and is capped with a thin 2 nm layer of Au to prevent oxidation. We then had to break vacuum and exchange sputtering targets. Ideally one would sputter the entire stack \textit{in situ}, however, we are limited to seven sputtering targets in our chamber. During the target exchange, which takes less than 10 minutes, the samples are contained in a bag filled with continuously flowing N$_2$ gas to limit contamination. After another bakeout, pump down, and liquid nitrogen cooling, we ion mill away the protective Au layer before continuing the sputtering process. All the ferromagnetic layers are then deposited \textit{in situ} in the following sequence: for the ``spin-singlet'' samples of Fig.~\ref{fig:SampleStructure}(a), Cu(4)/PMA-SAF/Cu(4)/Au(2); and for the ``spin-triplet'' samples of Fig.~\ref{fig:SampleStructure} (b) we sputter, Cu(2)/[Ni(1.6) or Py(1.25)]/Cu(4)/ PMA-SAF/Cu(4)/Ni(1.6)/Cu(7)/Au(2), where PMA-SAF=[Pd(0.9)/Co(0.3)]$_n$/Ru(0.95)/[Co(0.3)/Pd(0.9)]$_n$. Due to the crystal lattice mismatch between the fcc ferromagnetic materials and the bcc Nb we add a Cu(2) spacer before the (Py or Ni) $F^\prime$ layer. Meanwhile, between the $F^\prime$, $F$, and $F^{\prime \prime}$ layers, Cu(4) buffers are inserted to prevent them from coupling magnetically. Finally, the stack is capped with a thin layer of Cu and Au to prevent oxidation.

%For a wide figure spanning two columns use the figure* environment
\begin{figure*}
	\begin{center}
		\includegraphics[width=0.75\linewidth]{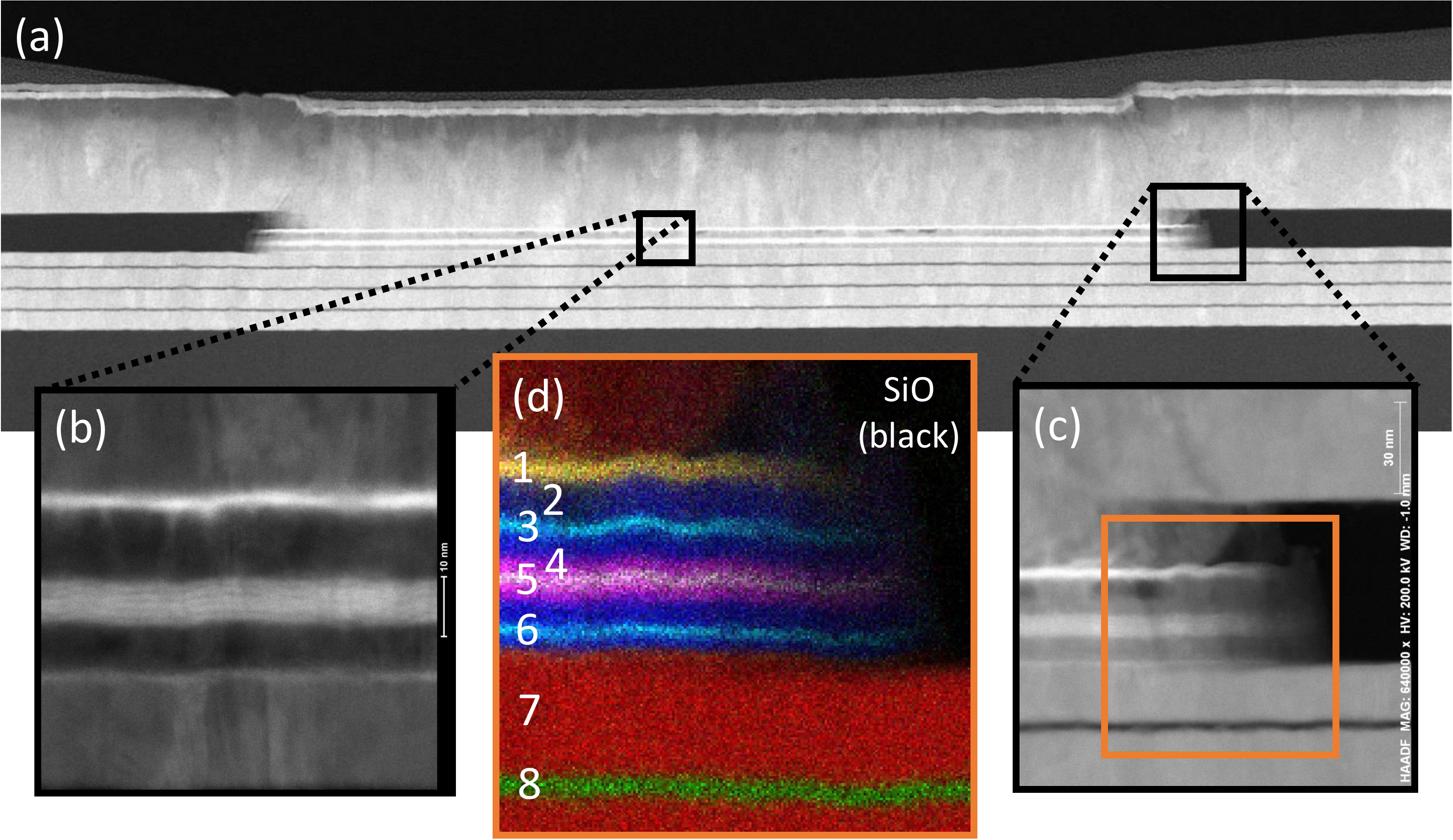}
	\end{center}
	\caption{\label{fig:SandiaSTEM} Vertical cross sections of the junctions described in Fig.~\ref{fig:SampleStructure} (b) prepared using a focused ion beam (FIB) were investigated by high-resolution scanning transmission electron microscopy (STEM) and energy dispersive x-ray spectroscopy (EDX). Panel (a) shows an STEM image of the full extent of the junction, including the smooth Nb/Al bottom electrode. Expanded views of the individual ferromagnetic layers are shown in panels (b) and (c) near the center and side of the junction, respectively. The EDX phase map shown in panel (d) corresponds to the area within the orange square in panel (c). The multivariate statistical analysis of the spectra from each individual pixel are color coded and numbered in the figure as follows: Au (yellow, 1), Cu (blue, 2), Ni + Fe (cyan, 3 and 6), Pd + Co (magenta, 4),  Co + Ru (white, 5), Nb (red, 7), Al (green, 8), SiO (black).}
\end{figure*}

The junctions were patterned by electron-beam lithography followed by ion milling in Argon. We use the negative e-beam resist ma-N2401 as the ion mill mask. The junctions are elliptical in shape with an aspect ratio of 2.5 and area of 0.5 $\mu$m$^2$, sufficiently small for the Py layers to be mostly single domain~\cite{Glick2016}. Elliptically-shaped junctions have the advantage that the Fraunhofer patterns follow an analytical formula while the (small) demagnetizing field is nearly uniform when the magnetization is uniform.

Outside the mask region, we ion milled from the capping layer through the $F^\prime$ layer, and nominally half-way into the underlying Cu spacer layer. After ion milling, a 50-nm-thick SiO layer was deposited by thermal evaporation to electrically isolate the junction and the bottom wiring layer from the top wiring layer.

Finally, the top Nb wiring layer was patterned using the same photolithography and lift-off process as the bottom leads. The surface is cleaned with oxygen plasma etching followed by \textit{in-situ} ion milling in which 1 nm of the top Au surface is removed immediately before sputtering. We deposited top leads of Nb(150 nm)/Au(10 nm), ending with the Au to prevent oxidation.

In order to verify the fabrication process, vertical cross sections of the junctions were investigated by high-resolution scanning transmission electron microscopy (STEM) and energy dispersive x-ray spectroscopy (EDX). The cross sections were prepared using a FEI Helios focused ion beam (FIB) with a Ga ion source, and transferred to a Ti grid for imaging in a FEI Titan G2 80-200 aberration-corrected STEM operated at 200kV and equipped with four silicon drift X-ray detectors.

The high-angle annular dark field STEM image in Figs.~\ref{fig:SandiaSTEM}(a)-(c) show cross sections of a junction with the same structure as Fig.~\ref{fig:SampleStructure} (b) with n = 3. The Nb/Al bottom electrode, seen at the bottom of Fig.~\ref{fig:SandiaSTEM}(a), is a smooth and continuous surface that provides a good growth template for the layers grown on top. Fourier transforms of regions of the high-resolution STEM image show that the Cu layer directly above grows with a [111] orientation on Nb [011]. Grains with favorable orientation relative to the beam direction show lattice fringes extending through the entire Cu/ferromagnetic layer/Cu thicknesses. In the top Cu(7 nm) layer there appear to be three to four isolated regions (width $\approx$ 5 nm) with lower Cu density than the rest of the layer (one such dark patch can be clearly seen in Fig.~\ref{fig:SandiaSTEM}(c)). The origin of those low-density Cu regions is unknown. The individual Pd and Co layers inside the SAF, shown in Fig.~\ref{fig:SandiaSTEM}(b), near the center of the junction, appear relatively smooth and continuous. Furthermore, the STEM image shows that the ion milling procedure used in the sample fabrication to define the junction area is accurately calibrated to mill down to the desired depth.

Identifying the elemental composition of the layers is achieved through EDX phase maps, created by performing a multivariate statistical analysis of the spectra from each individual pixel, and color-coding pixels containing the same spectral shape~\cite{Kotula2006b}. The phase map shown in Fig.~\ref{fig:SandiaSTEM}(d) corresponds to the area within the orange square in Fig.~\ref{fig:SandiaSTEM}(c). The Py layer (cyan, labeled as layer 6) is clearly uniform and continuous. The individual layers inside the SAF are not distinguishable due to their sub-nanometer thickness and the lower spatial resolution of EDX compared to STEM. However, we clearly show a difference between the component Pd/Co X-ray peaks from the outer regions of the SAF (magenta, labeled as layer 4) compared to its center, where component Ru/Co peaks are more promenent (white, labeled as layer 5). The Ni layer (cyan, labeled as layer 3), while continuous, has some observable roughness, consistent with the magnetic behavior discussed in the next section.
\section{\label{sec:level3}Transport Measurements and Analysis}
Each device was connected to the wire leads of a dip-stick probe with pressed indium solder. The samples were immersed in a liquid-He dewar outfitted with a Cryo-perm magnetic shield and placed inside a shielded room to reduce noise from external sources of electromagnetic radiation. The dipping probe is equipped with a superconducting solenoid used to apply uniform magnetic fields along the long-axis of the elliptical junctions. The current-voltage characteristics of the junctions were measured at 4.2 K in a four-terminal configuration.  The voltage across the junction is measured with a commercial Quantum Design rf SQUID in a self-balancing potentiometer comparator circuit, and the measurement current is provided by a battery-powered ultra-low noise programmable current source~\cite{Edmunds1980}. The rf SQUID comparator scheme has very low RMS voltage noise of only 6 pV when measuring over 10 power line cycles. Typical I-V curves have the expected behavior of overdamped Josephson junctions~\cite{Barone1982}. The critical current $I_{c}$ was extracted by fitting the I-V curves to a square root function of the form,
	\begin{equation}
	\label{eqn:SquareRootFunction}
	V= R_N \sqrt{I^2 - I_c^2}, \hspace*{0.1in} I \ge I_c,
	\end{equation}
where the sample resistance in the normal state $R_N$ was determined by the slope of the linear region of the I-V curve when $|I| \gg I_c$.

\begin{figure}
	\begin{center}
		\includegraphics[width=\linewidth]{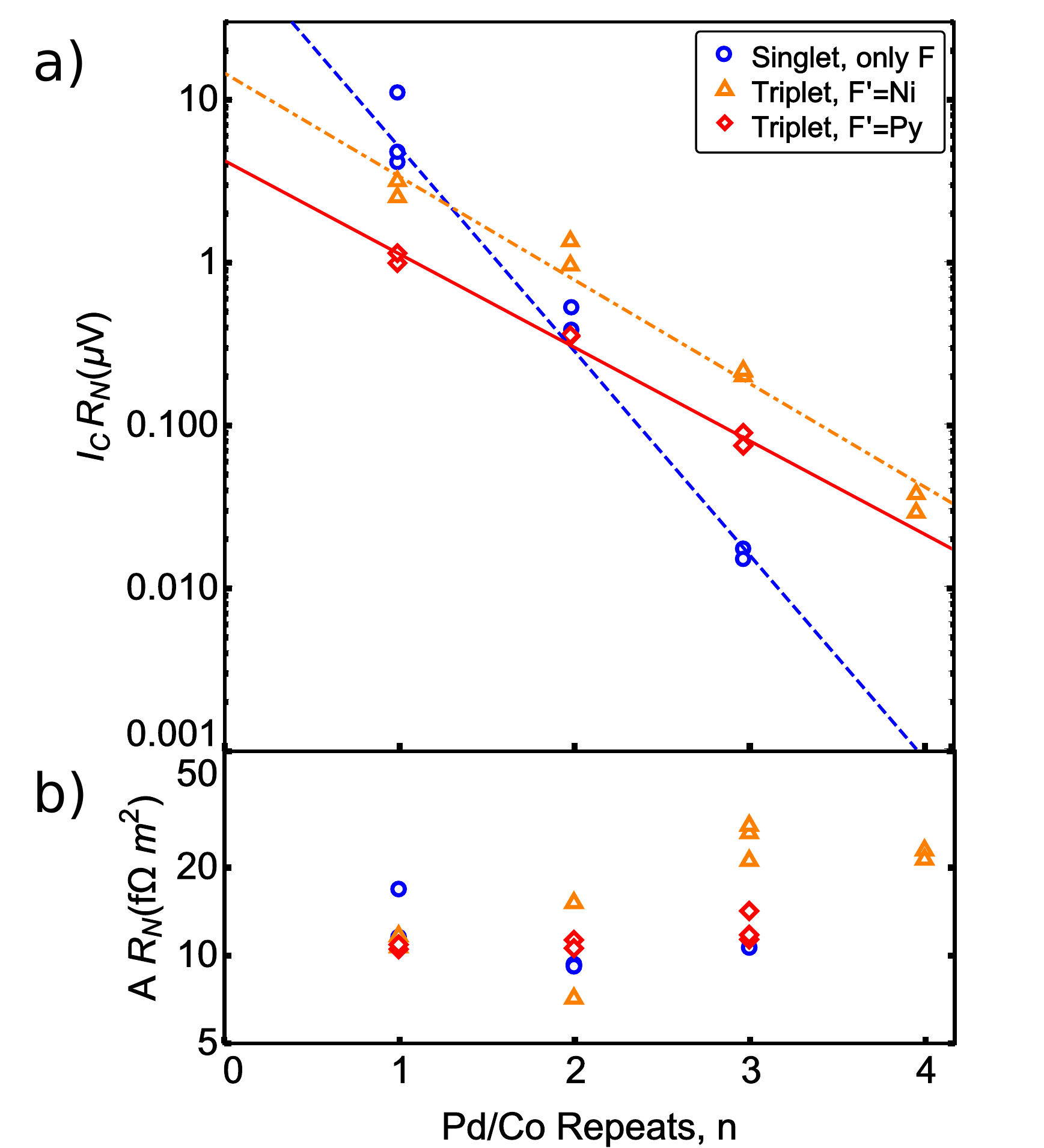}
	\end{center}
	\caption{\label{fig:IcRn_Vs_Thickness} (a) The maximum measured $I_c$ times $R_N$ is plotted vs. the number of Pd/Co repeats $n$ in the PMA SAF. The $I_c R_N$ of the spin-singlet samples (blue circles) decays more rapidly with increasing Pd/Co thickness than do the spin-triplet samples with $F^{\prime}$ = Ni (orange triangles) and those with $F^{\prime}$ = Py (red diamonds). The data are fit to the total number of Pd/Co bilayers according to Eq.~\ref{eq:ICRNexponentialdecay}, with the best-fit parameters shown in Table~\ref{table:IcRnFitParams}.
	(b) The area-resistance product of the junctions vs. $n$ does not appear to be correlated to the number of Pd/Co interfaces, and is on average 11.4 f$\Omega$m$^2$ across all the datasets, with the exception of the $F^\prime$=Ni chips with n = 3 and 4, which have slightly larger resistances.}
\end{figure}

When $I_c$ is less than a few $\mu$A, the I-V curves exhibit noticeable rounding due to thermal effects and instrumental noise.  Such rounding is accommodated by the theories of Ivanchenko and Zil'berman (IZ) and of Ambegaokar and Halperin~\cite{AmbegaokarHalperin1969, IvanchenkoZilberman1969}.  Fitting the I-V curves with the IZ function instead of the square-root function of Eqn.~\ref{eqn:SquareRootFunction} results in values of $I_c$ that are somewhat larger -- typically 30\% for I$_c \approx 1 \mu$A~\cite{Glick2016}. However, fitting every I-V curve with the IZ function is computation-intensive and not practical, so we used the simpler square-root fits for the Fraunhofer data shown in Figs.~\ref{fig:Singlet_Fraunhofers}-\ref{fig:Triplet_NiFe-Ni_Fraunhofers}.  For the summary shown in Fig.~\ref{fig:IcRn_Vs_Thickness}, we used the values obtained from fitting the IZ function to the data near the peaks of the individual Fraunhofer patterns.

Measurements of the area-resistance product in the normal state typically yield consistent values of $A R_N$, with a median value of 22.5 f$\Omega$-$m^2$, an indicator of the reproducible high quality interfaces. The $A R_N$ values for the full set of samples are shown in Fig.~\ref{fig:IcRn_Vs_Thickness}(b). The junction area typically varies by less than 10\% from the nominal value of 0.5 $\mu$m$^2$, and can be accurately extracted from the Fraunhofer pattern measurements discussed later. It is thought that $R_N$ is dominated by the interfacial resistance between the various layers. It is therefore noteworthy that, although these junctions contain many interfaces, the $A R_N$ products are (on average) only about twice those of similarly-sized junctions containing only a single ferromagnetic layer~\cite{Niedzielski2015, Glick2016}, which had 5-10 f$\Omega$-m$^2$. The junctions with $F^\prime=F^{\prime \prime}=$ Ni with $n$ = 3 or 4 Pd/Co repeats have slightly larger $A R_N$ than those with $n$ = 1 or 2. Otherwise, the average value of $A R_N$ does not appear to be correlated with $n$.

\subsection{Fraunhofer Patterns}
\begin{figure}
	\begin{center}
		\includegraphics[width=0.8\linewidth]{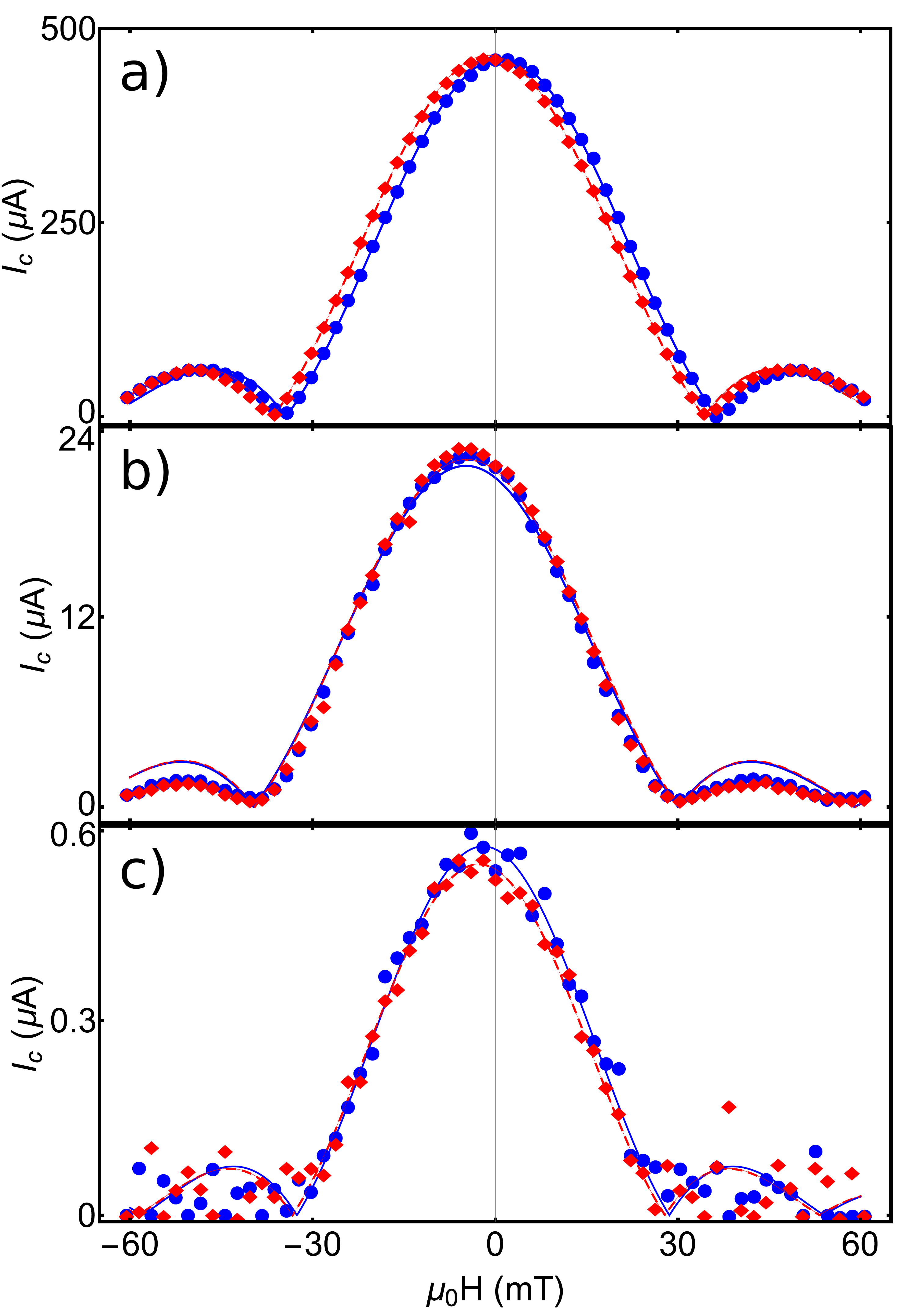}
	\end{center}
	\caption{\label{fig:Singlet_Fraunhofers} The critical current, $I_c$, is plotted versus the applied in-plane field $H$ for junctions with the structure shown in Fig.~\ref{fig:SampleStructure} (a) (without $F^\prime$ and $F^{\prime \prime}$). The supercurrent is carried primarily by spin-singlet pairs. Increasing the number of Pd/Co layer repeats: (a) $n=1$, (b) $n=2$, (c) $n=3$, causes $I_c$ to decay rapidly. Since the magnetization of the SAF is perpendicular to the plane there are only slight horizontal shifts in the Fraunhofer patterns and very little magnetic hysteresis. The corresponding fits to Eqn.~\ref{eqn:FraunhoferAiryFit} (lines) show excellent agreement for both the positive (red, dashed) and negative (blue) field sweep directions.}
\end{figure}

\begin{figure}
	\begin{center}
		\includegraphics[width=\linewidth]{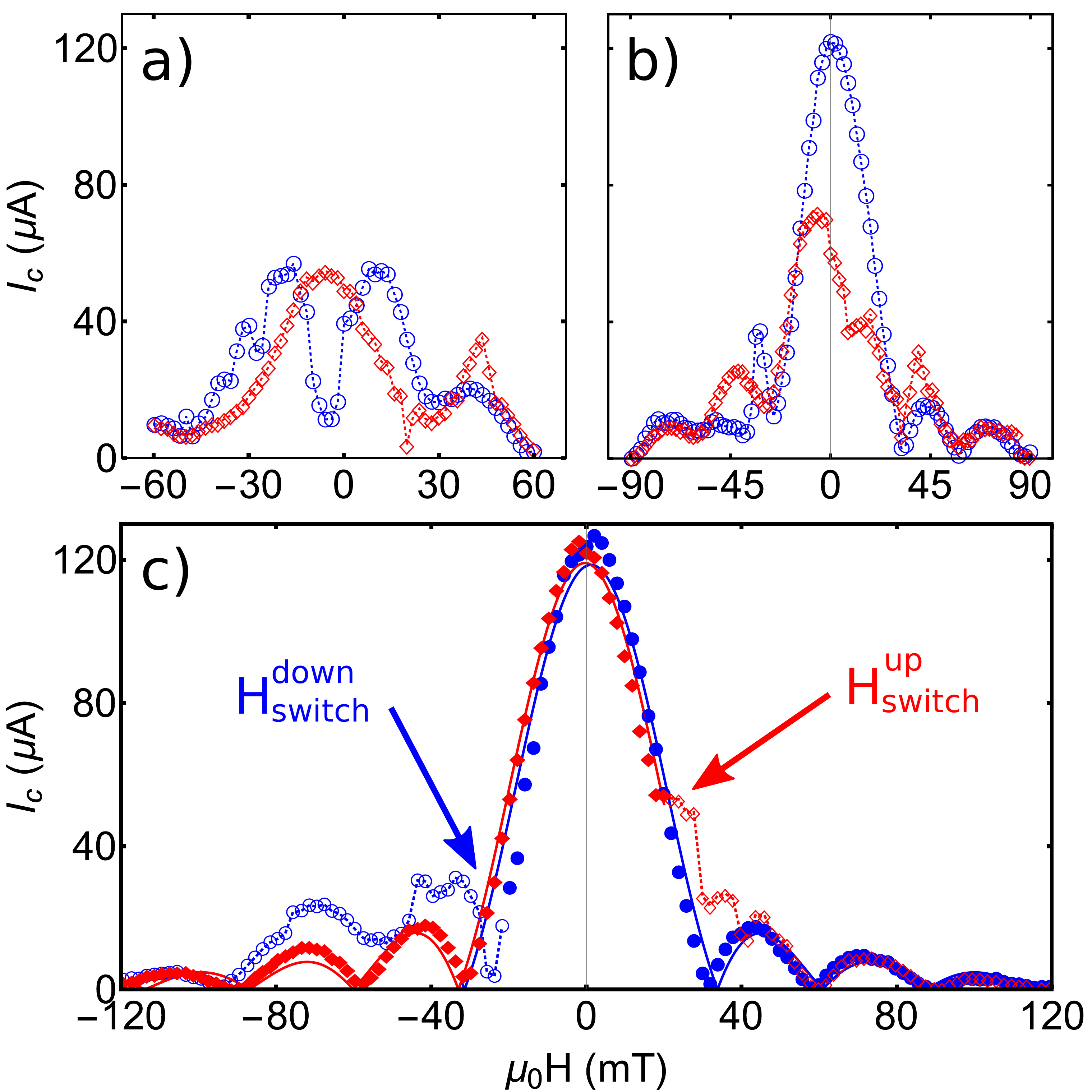}
	\end{center}
	\caption{\label{fig:Triplet_Ni-Ni_Fraunhofers} Critical current, $I_c$, vs. the applied in-plane field $H$, for a sample with $F^\prime$ = $F^{\prime \prime}$ = Ni (1.6 nm), and $n$=1, similar to Fig.~\ref{fig:SampleStructure} (b). If the initialization field is too small, e.g. (a) $H_{\mathrm{init}}$= 60 mT or (b) $H_{\mathrm{init}}$= 90 mT, and the magnetizations of the two Ni layers are not aligned, the ``Fraunhofer'' patterns are of poor quality. With an initialization field of 150 mT (c) the Ni layers are fully magnetized, and the data before $H_{\mathrm{switch}}$ (solid markers) show good agreement with the expected form for both the positive (red) and negative (blue) field sweep directions. Solid lines are fits to Eqn.~\ref{eqn:FraunhoferAiryFit}. Hence we initialized all our spin-triplet samples at 150 mT. In (c) the hollow markers represent the data points after $H_{\mathrm{switch}}$ and the dashed lines are only to guide the eye. The Ni layers, while amenable to large supercurrents, contain multiple magnetic domains and switch magnetization over a broad field range.}
\end{figure}

\begin{figure}
	\begin{center}
		\includegraphics[width=0.85\linewidth]{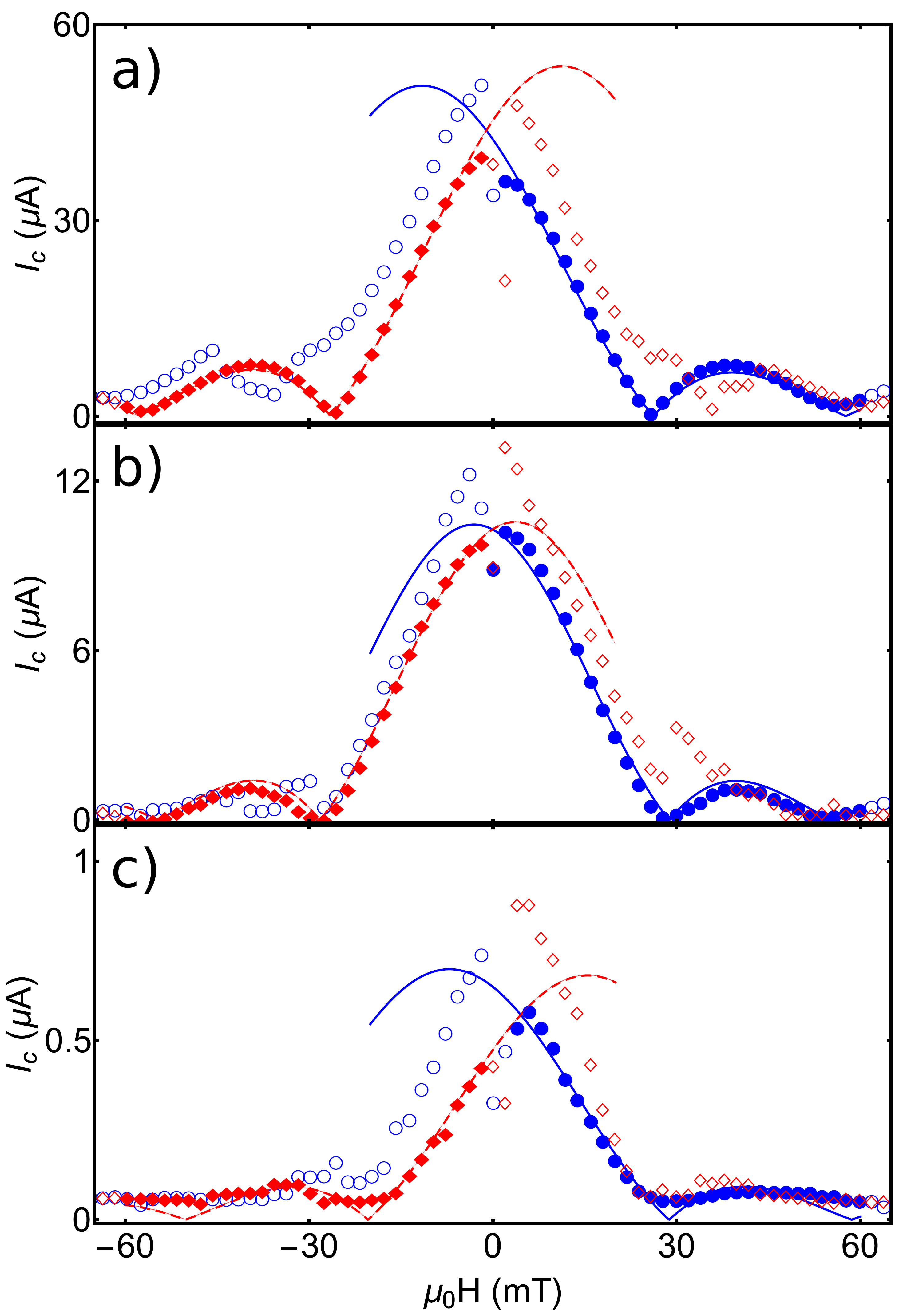}
	\end{center}
	\caption{\label{fig:Triplet_NiFe-Ni_Fraunhofers} $I_c$ is plotted vs. the applied in-plane field $H$, for junctions with the structure shown in Fig.~\ref{fig:SampleStructure} (b) with $F^\prime$ = Py, $F^{\prime \prime}$ = Ni. Increasing the number of repeats of Pd/Co: (a) n=1, (b) n=2, (c) n=3, causes I$_c$ to decay, but more slowly than without the $F^\prime$ and $F^{\prime \prime}$ layers (Fig.~\ref{fig:Singlet_Fraunhofers}). The horizontal shifts in the Fraunhofer patterns are indicative of the magnetic state of the in-plane ferromagnets. The data before $H_{\mathrm{switch}}$, the field at which the Py magnetization reverses direction (solid markers), and the corresponding fits to Eqn.~\ref{eqn:FraunhoferAiryFit} (lines) show excellent agreement for both the positive (red, dashed) and negative (blue) field sweep directions. The hollow circles are the corresponding data points after $H_{\mathrm{switch}}$. The Py typically switches abruptly at low fields ($<$ 2 mT), whereas the Ni is thought to contain multiple magnetic domains and gradually switches over a broad range of fields.}
\end{figure}

Measuring $I_c$ as a function of the applied magnetic field, we map out ``Fraunhofer'' diffraction patterns, shown in Figs.~\ref{fig:Singlet_Fraunhofers}-\ref{fig:Triplet_NiFe-Ni_Fraunhofers} for the three JJ types described in Fig.~\ref{fig:SampleStructure}. %To compare junctions with different cross-sectional areas we normalized our data by multiplying $I_c$ by $R_N$.

Fig.~\ref{fig:Singlet_Fraunhofers} shows data from three spin-singlet samples with $n$ = 1, 2, and 3.  Those data were acquired by applying a field of 60 mT then slowly ramping the field to -60 mT in steps of typically 2 mT (blue data points). We then repeated this procedure in the other field direction (red data points) and observed very little magnetic hysteresis since the F layer's magnetizations are aligned perpendicular to the applied field.

For the spin-triplet samples with the additional $F^\prime$ and $F^{\prime \prime}$ layers, we first measured the critical current near zero field to monitor the ``virgin'' or as-grown state of the various nanomagnets. Initially sweeping the field from ($\pm$ 60 mT) the critical current was typically small and any semblance of ``Fraunhofer patterns'' were rather irregular, as shown in Fig.~\ref{fig:Triplet_Ni-Ni_Fraunhofers} (a). Successively expanding the applied field sweep range ($\pm$60, $\pm$90, $\pm$120, $\pm$150 mT) to help align the $F^\prime$ and $F^{\prime \prime}$ layers resulted in significant improvements in the Fraunhofer quality and enhanced the peak value of I$_c$ (near zero field), as shown in Fig.~\ref{fig:Triplet_Ni-Ni_Fraunhofers}(b)-(c). After initializing the samples at $\pm$150 mT the peak value of I$_c$ appears to saturate, and the Fraunhofer pattern closely follows the theoretical curve described later in Eq.~\ref{eqn:FraunhoferAiryFit}. This detailed ``initialization'' behavior was reproducible on five separate junctions from various chips.  Therefore, for the data presented in Figs.~\ref{fig:Triplet_Ni-Ni_Fraunhofers}-\ref{fig:Triplet_NiFe-Ni_Fraunhofers}, we determined that a large initialization field of 150 mT was required to help set the initial orientation of the Ni layer(s). Ideally, an even larger initialization field would be beneficial to fully magnetize the Ni~\cite{Baek_privatecommunication}, however, too large a field might disturb the magnetic properties of the PMA SAF. Based on the results of Figs.~\ref{fig:M-Hloop_PdCo_10Repeats}-\ref{fig:M-Hloop_PdCo_SAF}, as a precautionary measure any external magnetic fields were kept at 150 mT or below.

After the initialization procedure, we removed any flux trapped in the junction or in the Nb leads by lifting the dip-stick probe slightly in the Dewar until the sample lay just above the liquid Helium bath. After reinserting the sample into the liquid Helium we next applied a field of 90 mT and slowly ramped the field to -90 mT, in steps of 2 mT. Finally, the field was slowly swept in the opposite direction, observing any hysteretic effects from the in-plane ferromagnets.

Fraunhofer pattern measurements such as these contain information about the magnetic state of the in-plane ferromagnetic layers, the behavior of the critical current, and the dimensions of the junction. For elliptical junctions the functional form of the Fraunhofer pattern is an Airy function~\cite{Barone1982},
\begin{equation}
\label{eqn:FraunhoferAiryFit}
I_{c}=I_{c0} \left| 2 J_{1} \left( \pi \Phi / \Phi_{0} \right) / \left( \pi \Phi / \Phi_0 \right) \right|,
\end{equation}
where $J_1$ is a Bessel function of the first kind, $I_{c0}$ is the maximum critical current, $\Phi$ is the total magnetic flux due to both the external field and the magnetic layers in the junction, and $\Phi_0 = h/2e$ is the flux quantum. Since the magnetization of the F-layer is parallel to the current flow, it does not contribute to $\Phi$. The in-plane magnetizations of $F^\prime$ and $F^{\prime \prime}$ do contribute, however, and cause shifts in the Fraunhofer pattern along the field axis~\cite{Ryazanov1999, Khaire2009, Baek2014,Niedzielski2015, Glick2016, Niedzielski2017}.  Hence the horizontal Fraunhofer pattern shifts will differ for the three types of samples outlined in Fig.~\ref{fig:SampleStructure}. For samples without the $F^\prime$ and $F^{\prime \prime}$ layers there should be very little shift, resulting only from any canting of the [Pd/Co] perpendicular magnetization into the plane. That expectation is born out by the data shown in Fig.~\ref{fig:Singlet_Fraunhofers}. For samples with the $F^\prime$ and $F^{\prime \prime}$ layers the shifts will be more pronounced. The shifts should be largest when the magnetizations of $F^{\prime}$ and $F^{\prime \prime}$ are parallel and smaller when they are antiparallel~\cite{Niedzielski2017}.

The complex nature of these Fraunhofer pattern shifts, combined with the possibility that the Ni layers do not switch abruptly or behave as a single magnetic domain, make a comprehensive analysis difficult. For the samples with $F^\prime$ = $F^{\prime \prime}$, the two Ni layers may not switch at the same field due to the upward propagating surface roughness, making it impractical to fit all the data points in the Fraunhofer patterns shown in Figs. ~\ref{fig:Triplet_Ni-Ni_Fraunhofers} and \ref{fig:Triplet_NiFe-Ni_Fraunhofers}. Therefore we only attempt to fit the clean sections of the Fraunhofer patterns before the first magnetic switching event occurs. We fit Eqn.~\ref{eqn:FraunhoferAiryFit} to the data starting from the initialization field to $H_{\mathrm{switch}}$. The free parameters in the fit are $I_{c0}$, the junction width transverse to the field direction, and the field shift of the central peak. In Figs.~\ref{fig:Singlet_Fraunhofers} - \ref{fig:Triplet_NiFe-Ni_Fraunhofers} the corresponding fits show excellent agreement with the data, for the positive (red) and negative (blue) sweep directions. In Figs.~\ref{fig:Triplet_Ni-Ni_Fraunhofers}(c) and \ref{fig:Triplet_NiFe-Ni_Fraunhofers} the hollow data points denote the data after $H_{\mathrm{switch}}$. Those data show that the reversal of the Ni magnetization occurs over a range of fields, consistent with the behavior of Ni seen in previous work by us and others~\cite{Baek2014, Niedzielski2017}. Most of the junctions display full magnetic remanence, continuing to follow Eqn.~\ref{eqn:FraunhoferAiryFit} though zero applied field before $H_{\mathrm{switch}}$.

\subsection{Critical Current vs Thickness}
In Fig.~\ref{fig:IcRn_Vs_Thickness}(a) we plot the maximum measured $I_c$ times $R_N$ on a log scale versus the number of Pd/Co repeats $n$ on either side of the Ru spacer. In Fig.\ref{fig:IcRn_Vs_Thickness} (b) we plot the area-resistance product for the entire data set.  The $I_c R_N$ products of the spin-singlet samples (blue circles) decay much more rapidly with increasing $n$ than do the spin-triplet samples with $F^{\prime}$ = Ni (orange triangles) and those with $F^{\prime}$ = Py (red diamonds).  The three data sets are fit to the total number of bilayers (2$n$) with a simple exponential decay,
\begin{equation}
\label{eq:ICRNexponentialdecay}
I_c R_N (n) = A_0*e^{-(2 n/ \bar{n})}.
\end{equation}
The best-fit parameters are listed in Table~\ref{table:IcRnFitParams}.  The decay length expressed as a number of [Co/Pd] bilayers is 1.38$\pm$0.07 and 1.53$\pm$0.07 for the spin-triplet samples, whereas it is only 0.70$\pm$0.04 for the spin-singlet samples.  That observation is the main result of this work.
\setlength{\tabcolsep}{8pt} % default is 6pt
\renewcommand{\arraystretch}{1.5} % default is 1.0
\begin{table}
\centering
\caption{\label{table:IcRnFitParams} Fitting the data in Fig.~\ref{fig:IcRn_Vs_Thickness} to Eqn.~\ref{eq:ICRNexponentialdecay} yields the following best-fit parameters:}
	\begin{tabular}{lcc@{\hskip 0.1in}}
	{Sample Set}         &  {A$_0$ ($\mu$V)}  &  $\bar{n}$  \\ \hline\hline %\noalign{\smallskip} 
	Spin-singlet, only F & 89 $\pm$ 28        & 0.70 $\pm$ 0.04\\ \hline %\noalign{\smallskip}
	Spin-triplet, $F^\prime$=$F^{\prime \prime}$= Ni & \hspace{-0.28cm} 14 $\pm$ 3 & 1.38 $\pm$ 0.07\\ \hline %\noalign{\smallskip}
	Spin-triplet, $F^\prime$=Py, $F^{\prime \prime}$=Ni & 4.1 $\pm$ 0.6 & 1.53 $\pm$ 0.07\\ \hline %\noalign{\smallskip}
\end{tabular}
\setlength{\tabcolsep}{12pt}
\end{table}

We believe that the spin-scattering asymmetry in [Co/Pd] underlies the fact that the decay of the spin-triplet supercurrent is only a factor of two less steep than the decay of the spin-singlet supercurrent. To see why, it is instructive to compare the data in Fig.\ref{fig:IcRn_Vs_Thickness}(a) with data from the only other study of spin-triplet Josephson junctions containing PMA layers, namely ref. \onlinecite{Gingrich2012}. In that work the central F layer consisted of a [Ni/Co] multilayer with strong PMA, but not a SAF. In those junctions the decay of the spin-triplet samples was much slower than the decay of the control samples that did not contain the $F^\prime$ and $F^{\prime \prime}$ layers. (The ratio of the spin-triplet to spin-single decay lengths in that work was about 4.5, whereas it is only about 2.1 in the present work.) Later, our group measured a series of junctions containing [Ni/Co] PMA SAFs, and found not only that they carried much smaller critical currents, but that the decay with the number of repeats was also much steeper than in the non-SAF [Ni/Co] junctions~\cite{Korucu-unpublished}. To explain the sharp decay of the spin-triplet supercurrent in the [Ni/Co] SAFs, we propose the following explanation. It is known from Giant Magnetoresistance (GMR) studies~\cite{Nguyen2010} that the [Ni/Co] interface has strong spin-scattering asymmetry -- i.e. minority-band electrons are scattered more strongly than majority-band electrons. Hence in a [Ni/Co]$_n$ multilayer, the current will become more strongly spin-polarized as the number of repeats $n$ increases. In a [Ni/Co] SAF, the majority electrons in one half of the SAF become minority electrons in the other half and vice versa. Hence all electrons passing through the SAF are strongly scattered at some point. This effect increases with $n$, hence causing a steep decay of critical current as a function of $n$. We note that this mechanism affects both spin-singlet and spin-triplet supercurrents; but the former already face a steep decay due to the standard S/F physics discussed in the introduction, whereas for the spin-triplet supercurrent the presence of the SAF becomes the dominant decay mechanism.

Our decision to use [Pd/Co] multilayers was a direct result of the discussion above.  Unfortunately, the degree of spin-scattering asymmetry at the [Pd/Co] interface has not been measured, as far as we know. (Our own attempts to do so using GMR techniques were thwarted by our inability to achieve reproducible in-plane magnetic states for any thickness combination in the [Pd/Co] system.)  From comparing the results shown in Fig.~\ref{fig:IcRn_Vs_Thickness}(a) with our unpublished data on junctions containing [Ni/Co] PMA SAFs, we infer that the spin-scattering asymmetry in [Pd/Co] is weaker than in [Ni/Co].

An alternative explanation for the unusually steep decay of the spin-triplet supercurrent in our samples is the strong spin-orbit interaction in the Co/Pd system~\cite{Pollard2017}.  It should be possible to distinguish between these two explanations by comparing the decay of the supercurrent in spin-triplet JJs containing [Pd/Co] plain multilayers in the center, with the decay we observe in our JJs containing [Pd/Co] multilayer SAFs.  We intend to carry out such a study in the near future.

The next question to address is, at what value of $n$ can we be certain that the spin-triplet component of the supercurrent in the spin-triplet samples is the dominant contribution.  A straightforward interpretation of the data shown in Fig.\ref{fig:IcRn_Vs_Thickness}(a) might lead one to conclude that the point where the spin-triplet curves cross the spin-singlet curve is the point where the spin-singlet and spin-triplet contributions to the supercurrent are equal in the spin-triplet samples.  Those crossings occur approximately at $n=1$ for the Ni-Ni samples and at $n=2$ for the Ni-Py samples.  But that interpretation is wrong.  The magnitude of the spin-singlet supercurrent in the spin-triplet samples is surely far less than the supercurrent we measure in the spin-singlet samples, because the spin-singlet supercurrent will be further suppressed when it has to pass through the additional $F^{\prime}$ and $F^{\prime \prime}$ layers. That suppression would effectively shift the entire spin-singlet curve down vertically, thus the blue data set can only be considered as a generous upper bound on the magnitude of the spin-singlet supercurrent that can pass through the spin-triplet samples. We do not know how large that suppression is, but we can guess that the suppression is roughly three times greater in the Py-Ni samples than in the Ni-Ni samples, from the vertical offset between the two spin-triplet curves.

\section{Conclusion}
In conclusion, we have measured the critical current in Josephson junctions of the form $S/F^{\prime}/N/F/N/F^{\prime \prime}/S$, where F is a synthetic antiferromagnets consisting of [Pd/Co] multilayers with perpendicular anisotropy.  The critical currents in those junctions decay less steeply with the number of [Pd/Co] bilayers than in junctions without the $F^{\prime}$ and $F^{\prime \prime}$ layers. That result represents strong evidence that the 3-layer junctions carry spin-triplet supercurrent. The central SAF, which achieves the long-range spin-triplet conversion, also serves to reduce stray magnetic fields detrimental to the other ferromagnetic layers. Furthermore, by choosing $F^{\prime}$ to be a soft magnetic material such as Permalloy, while $F^{\prime \prime}$ is a hard magnetic material such as Ni, the relative magnetization directions of those two layers can be controlled. Such junctions hold promise for future experiments where the ground-state phase difference across the junction can be controlled by changing the magnetic configuration.

Acknowledgements: We thank E. Gingrich, A. Herr, D. Miller, O. Namann, N. Rizzo, and M. Schneider for helpful discussions, B. Bi for help with fabrication using the Keck Microfabrication Facility and L. Lowery for FIB sample preparation. This research is supported by the Office of the Director of National Intelligence (ODNI), Intelligence Advanced Research Projects Activity (IARPA), via U.S. Army Research Office contract W911NF-14-C-0115. The views and conclusions contained herein are those of the authors and should not be interpreted as necessarily representing the official policies or endorsements, either expressed or implied, of the ODNI, IARPA, or the U.S. Government. Sandia National Laboratories is a multi-mission laboratory managed and operated by National Technology and Engineering Solutions of Sandia LLC, a wholly owned subsidiary of Honeywell International Inc. for the U.S. Department of Energy’s National Nuclear Security Administration under contract DE-NA0003525.
\nocite{*}
\bibliography{Glick_Triplet_2017}

\begin{thebibliography}{72}%
\makeatletter
\providecommand \@ifxundefined [1]{%
 \@ifx{#1\undefined}
}%
\providecommand \@ifnum [1]{%
 \ifnum #1\expandafter \@firstoftwo
 \else \expandafter \@secondoftwo
 \fi
}%
\providecommand \@ifx [1]{%
 \ifx #1\expandafter \@firstoftwo
 \else \expandafter \@secondoftwo
 \fi
}%
\providecommand \natexlab [1]{#1}%
\providecommand \enquote  [1]{``#1''}%
\providecommand \bibnamefont  [1]{#1}%
\providecommand \bibfnamefont [1]{#1}%
\providecommand \citenamefont [1]{#1}%
\providecommand \href@noop [0]{\@secondoftwo}%
\providecommand \href [0]{\begingroup \@sanitize@url \@href}%
\providecommand \@href[1]{\@@startlink{#1}\@@href}%
\providecommand \@@href[1]{\endgroup#1\@@endlink}%
\providecommand \@sanitize@url [0]{\catcode `\\12\catcode `\$12\catcode
  `\&12\catcode `\#12\catcode `\^12\catcode `\_12\catcode `\%12\relax}%
\providecommand \@@startlink[1]{}%
\providecommand \@@endlink[0]{}%
\providecommand \url  [0]{\begingroup\@sanitize@url \@url }%
\providecommand \@url [1]{\endgroup\@href {#1}{\urlprefix }}%
\providecommand \urlprefix  [0]{URL }%
\providecommand \Eprint [0]{\href }%
\providecommand \doibase [0]{http://dx.doi.org/}%
\providecommand \selectlanguage [0]{\@gobble}%
\providecommand \bibinfo  [0]{\@secondoftwo}%
\providecommand \bibfield  [0]{\@secondoftwo}%
\providecommand \translation [1]{[#1]}%
\providecommand \BibitemOpen [0]{}%
\providecommand \bibitemStop [0]{}%
\providecommand \bibitemNoStop [0]{.\EOS\space}%
\providecommand \EOS [0]{\spacefactor3000\relax}%
\providecommand \BibitemShut  [1]{\csname bibitem#1\endcsname}%
\let\auto@bib@innerbib\@empty
%</preamble>
\bibitem [{\citenamefont {Buzdin}(2005)}]{Buzdin_SFReview2005}%
  \BibitemOpen
  \bibfield  {author} {\bibinfo {author} {\bibfnamefont {A.~I.}\ \bibnamefont
  {Buzdin}},\ }\href {\doibase 10.1103/RevModPhys.77.935} {\bibfield  {journal}
  {\bibinfo  {journal} {Rev. Mod. Phys.}\ }\textbf {\bibinfo {volume} {77}},\
  \bibinfo {pages} {935} (\bibinfo {year} {2005})}\BibitemShut {NoStop}%
\bibitem [{\citenamefont {Fulde}\ and\ \citenamefont
  {Ferrell}(1964)}]{Fulde1964}%
  \BibitemOpen
  \bibfield  {author} {\bibinfo {author} {\bibfnamefont {P.}~\bibnamefont
  {Fulde}}\ and\ \bibinfo {author} {\bibfnamefont {R.~A.}\ \bibnamefont
  {Ferrell}},\ }\href {\doibase 10.1103/PhysRev.135.A550} {\bibfield  {journal}
  {\bibinfo  {journal} {Phys. Rev.}\ }\textbf {\bibinfo {volume} {135}},\
  \bibinfo {pages} {A550} (\bibinfo {year} {1964})}\BibitemShut {NoStop}%
\bibitem [{\citenamefont {Larkin}\ and\ \citenamefont
  {Ovchinnikov}(1964)}]{Larkin1964}%
  \BibitemOpen
  \bibfield  {author} {\bibinfo {author} {\bibfnamefont {A.~I.}\ \bibnamefont
  {Larkin}}\ and\ \bibinfo {author} {\bibfnamefont {Y.~N.}\ \bibnamefont
  {Ovchinnikov}},\ }\href@noop {} {\bibfield  {journal} {\bibinfo  {journal}
  {Zh. Eksp. Teor. Fiz}\ }\textbf {\bibinfo {volume} {47}},\ \bibinfo {pages}
  {1136} (\bibinfo {year} {1964})},\ \bibinfo {note} {[\textit{Soviet Physics
  J. Exp. Theor. Phys.}, \textbf{20}, 762, (1965)]}\BibitemShut {NoStop}%
\bibitem [{\citenamefont {Buzdin}\ \emph {et~al.}(1982)\citenamefont {Buzdin},
  \citenamefont {Bulaevskii},\ and\ \citenamefont {Panyukov}}]{Buzdin1982}%
  \BibitemOpen
  \bibfield  {author} {\bibinfo {author} {\bibfnamefont {A.~I.}\ \bibnamefont
  {Buzdin}}, \bibinfo {author} {\bibfnamefont {L.~N.}\ \bibnamefont
  {Bulaevskii}}, \ and\ \bibinfo {author} {\bibfnamefont {S.~V.}\ \bibnamefont
  {Panyukov}},\ }\href@noop {} {\bibfield  {journal} {\bibinfo  {journal}
  {Pis'ma Zh. Eksp. Teor. Fiz.}\ }\textbf {\bibinfo {volume} {35}},\ \bibinfo
  {pages} {147} (\bibinfo {year} {1982})},\ \bibinfo {note} {[\textit{J. Exp.
  Theor. Phys. Lett.}, \textbf{35}, 4, 20 (1982)]}\BibitemShut {NoStop}%
\bibitem [{\citenamefont {Demler}\ \emph {et~al.}(1997)\citenamefont {Demler},
  \citenamefont {Arnold},\ and\ \citenamefont {Beasley}}]{Demler1997}%
  \BibitemOpen
  \bibfield  {author} {\bibinfo {author} {\bibfnamefont {E.~A.}\ \bibnamefont
  {Demler}}, \bibinfo {author} {\bibfnamefont {G.~B.}\ \bibnamefont {Arnold}},
  \ and\ \bibinfo {author} {\bibfnamefont {M.~R.}\ \bibnamefont {Beasley}},\
  }\href {\doibase 10.1103/PhysRevB.55.15174} {\bibfield  {journal} {\bibinfo
  {journal} {Phys. Rev. B}\ }\textbf {\bibinfo {volume} {55}},\ \bibinfo
  {pages} {15174} (\bibinfo {year} {1997})}\BibitemShut {NoStop}%
\bibitem [{\citenamefont {Ryazanov}\ \emph {et~al.}(2001)\citenamefont
  {Ryazanov}, \citenamefont {Oboznov}, \citenamefont {Rusanov}, \citenamefont
  {Veretennikov}, \citenamefont {Golubov},\ and\ \citenamefont
  {Aarts}}]{Ryazanov2001}%
  \BibitemOpen
  \bibfield  {author} {\bibinfo {author} {\bibfnamefont {V.~V.}\ \bibnamefont
  {Ryazanov}}, \bibinfo {author} {\bibfnamefont {V.~A.}\ \bibnamefont
  {Oboznov}}, \bibinfo {author} {\bibfnamefont {A.~Y.}\ \bibnamefont
  {Rusanov}}, \bibinfo {author} {\bibfnamefont {A.~V.}\ \bibnamefont
  {Veretennikov}}, \bibinfo {author} {\bibfnamefont {A.~A.}\ \bibnamefont
  {Golubov}}, \ and\ \bibinfo {author} {\bibfnamefont {J.}~\bibnamefont
  {Aarts}},\ }\href@noop {} {\bibfield  {journal} {\bibinfo  {journal} {Phys.
  Rev. Lett.}\ }\textbf {\bibinfo {volume} {86}},\ \bibinfo {pages} {2427}
  (\bibinfo {year} {2001})}\BibitemShut {NoStop}%
\bibitem [{\citenamefont {Kontos}\ \emph {et~al.}(2002)\citenamefont {Kontos},
  \citenamefont {Aprili}, \citenamefont {Lesueur}, \citenamefont {Gen{\^e}t},
  \citenamefont {Stephanidis},\ and\ \citenamefont {Boursier}}]{Kontos2002}%
  \BibitemOpen
  \bibfield  {author} {\bibinfo {author} {\bibfnamefont {T.}~\bibnamefont
  {Kontos}}, \bibinfo {author} {\bibfnamefont {M.}~\bibnamefont {Aprili}},
  \bibinfo {author} {\bibfnamefont {J.}~\bibnamefont {Lesueur}}, \bibinfo
  {author} {\bibfnamefont {F.}~\bibnamefont {Gen{\^e}t}}, \bibinfo {author}
  {\bibfnamefont {B.}~\bibnamefont {Stephanidis}}, \ and\ \bibinfo {author}
  {\bibfnamefont {R.}~\bibnamefont {Boursier}},\ }\href {\doibase
  10.1103/PhysRevLett.89.137007} {\bibfield  {journal} {\bibinfo  {journal}
  {Phys. Rev. Lett.}\ }\textbf {\bibinfo {volume} {89}},\ \bibinfo {pages}
  {137007} (\bibinfo {year} {2002})}\BibitemShut {NoStop}%
\bibitem [{\citenamefont {Blum}\ \emph {et~al.}(2002)\citenamefont {Blum},
  \citenamefont {Tsukernik}, \citenamefont {Karpovski},\ and\ \citenamefont
  {Palevski}}]{Blum2002}%
  \BibitemOpen
  \bibfield  {author} {\bibinfo {author} {\bibfnamefont {Y.}~\bibnamefont
  {Blum}}, \bibinfo {author} {\bibfnamefont {A.}~\bibnamefont {Tsukernik}},
  \bibinfo {author} {\bibfnamefont {M.}~\bibnamefont {Karpovski}}, \ and\
  \bibinfo {author} {\bibfnamefont {A.}~\bibnamefont {Palevski}},\ }\href
  {\doibase 10.1103/PhysRevLett.89.187004} {\bibfield  {journal} {\bibinfo
  {journal} {Phys. Rev. Lett.}\ }\textbf {\bibinfo {volume} {89}},\ \bibinfo
  {pages} {187004} (\bibinfo {year} {2002})}\BibitemShut {NoStop}%
\bibitem [{\citenamefont {Sellier}\ \emph {et~al.}(2003)\citenamefont
  {Sellier}, \citenamefont {Baraduc}, \citenamefont {Lefloch},\ and\
  \citenamefont {Calemczuk}}]{Sellier2003}%
  \BibitemOpen
  \bibfield  {author} {\bibinfo {author} {\bibfnamefont {H.}~\bibnamefont
  {Sellier}}, \bibinfo {author} {\bibfnamefont {C.}~\bibnamefont {Baraduc}},
  \bibinfo {author} {\bibfnamefont {F.}~\bibnamefont {Lefloch}}, \ and\
  \bibinfo {author} {\bibfnamefont {R.}~\bibnamefont {Calemczuk}},\ }\href
  {\doibase 10.1103/PhysRevB.68.054531} {\bibfield  {journal} {\bibinfo
  {journal} {Phys. Rev. B}\ }\textbf {\bibinfo {volume} {68}},\ \bibinfo
  {pages} {054531} (\bibinfo {year} {2003})}\BibitemShut {NoStop}%
\bibitem [{\citenamefont {Shelukhin}\ \emph {et~al.}(2006)\citenamefont
  {Shelukhin}, \citenamefont {Tsukernik}, \citenamefont {Karpovski},
  \citenamefont {Blum}, \citenamefont {Efetov}, \citenamefont {Volkov},
  \citenamefont {Champel}, \citenamefont {Eschrig}, \citenamefont
  {L{\"o}fwander}, \citenamefont {Sch{\"o}n},\ and\ \citenamefont
  {Palevski}}]{Shelukhin2006}%
  \BibitemOpen
  \bibfield  {author} {\bibinfo {author} {\bibfnamefont {V.}~\bibnamefont
  {Shelukhin}}, \bibinfo {author} {\bibfnamefont {A.}~\bibnamefont
  {Tsukernik}}, \bibinfo {author} {\bibfnamefont {M.}~\bibnamefont
  {Karpovski}}, \bibinfo {author} {\bibfnamefont {Y.}~\bibnamefont {Blum}},
  \bibinfo {author} {\bibfnamefont {K.~B.}\ \bibnamefont {Efetov}}, \bibinfo
  {author} {\bibfnamefont {A.~F.}\ \bibnamefont {Volkov}}, \bibinfo {author}
  {\bibfnamefont {T.}~\bibnamefont {Champel}}, \bibinfo {author} {\bibfnamefont
  {M.}~\bibnamefont {Eschrig}}, \bibinfo {author} {\bibfnamefont
  {T.}~\bibnamefont {L{\"o}fwander}}, \bibinfo {author} {\bibfnamefont
  {G.}~\bibnamefont {Sch{\"o}n}}, \ and\ \bibinfo {author} {\bibfnamefont
  {A.}~\bibnamefont {Palevski}},\ }\href {\doibase 10.1103/PhysRevB.73.174506}
  {\bibfield  {journal} {\bibinfo  {journal} {Phys. Rev. B}\ }\textbf {\bibinfo
  {volume} {73}},\ \bibinfo {pages} {174506} (\bibinfo {year}
  {2006})}\BibitemShut {NoStop}%
\bibitem [{\citenamefont {Weides}\ \emph {et~al.}(2006)\citenamefont {Weides},
  \citenamefont {Kemmler}, \citenamefont {Goldobin}, \citenamefont {Koelle},
  \citenamefont {Kleiner}, \citenamefont {Kohlstedt},\ and\ \citenamefont
  {Buzdin}}]{Weides2006}%
  \BibitemOpen
  \bibfield  {author} {\bibinfo {author} {\bibfnamefont {M.}~\bibnamefont
  {Weides}}, \bibinfo {author} {\bibfnamefont {M.}~\bibnamefont {Kemmler}},
  \bibinfo {author} {\bibfnamefont {E.}~\bibnamefont {Goldobin}}, \bibinfo
  {author} {\bibfnamefont {D.}~\bibnamefont {Koelle}}, \bibinfo {author}
  {\bibfnamefont {R.}~\bibnamefont {Kleiner}}, \bibinfo {author} {\bibfnamefont
  {H.}~\bibnamefont {Kohlstedt}}, \ and\ \bibinfo {author} {\bibfnamefont
  {A.}~\bibnamefont {Buzdin}},\ }\href@noop {} {\bibfield  {journal} {\bibinfo
  {journal} {Appl. Phys. Lett.}\ }\textbf {\bibinfo {volume} {89}},\ \bibinfo
  {pages} {122511} (\bibinfo {year} {2006})}\BibitemShut {NoStop}%
\bibitem [{\citenamefont {Robinson}\ \emph {et~al.}(2006)\citenamefont
  {Robinson}, \citenamefont {Piano}, \citenamefont {Burnell}, \citenamefont
  {Bell},\ and\ \citenamefont {Blamire}}]{Robinson2006}%
  \BibitemOpen
  \bibfield  {author} {\bibinfo {author} {\bibfnamefont {J.~W.~A.}\
  \bibnamefont {Robinson}}, \bibinfo {author} {\bibfnamefont {S.}~\bibnamefont
  {Piano}}, \bibinfo {author} {\bibfnamefont {G.}~\bibnamefont {Burnell}},
  \bibinfo {author} {\bibfnamefont {C.}~\bibnamefont {Bell}}, \ and\ \bibinfo
  {author} {\bibfnamefont {M.~G.}\ \bibnamefont {Blamire}},\ }\href {\doibase
  10.1103/PhysRevLett.97.177003} {\bibfield  {journal} {\bibinfo  {journal}
  {Phys. Rev. Lett.}\ }\textbf {\bibinfo {volume} {97}},\ \bibinfo {pages}
  {177003} (\bibinfo {year} {2006})}\BibitemShut {NoStop}%
\bibitem [{\citenamefont {Robinson}\ \emph {et~al.}(2007)\citenamefont
  {Robinson}, \citenamefont {Piano}, \citenamefont {Burnell}, \citenamefont
  {Bell},\ and\ \citenamefont {Blamire}}]{Robinson2007}%
  \BibitemOpen
  \bibfield  {author} {\bibinfo {author} {\bibfnamefont {J.~W.~A.}\
  \bibnamefont {Robinson}}, \bibinfo {author} {\bibfnamefont {S.}~\bibnamefont
  {Piano}}, \bibinfo {author} {\bibfnamefont {G.}~\bibnamefont {Burnell}},
  \bibinfo {author} {\bibfnamefont {C.}~\bibnamefont {Bell}}, \ and\ \bibinfo
  {author} {\bibfnamefont {M.~G.}\ \bibnamefont {Blamire}},\ }\href {\doibase
  10.1103/PhysRevB.76.094522} {\bibfield  {journal} {\bibinfo  {journal} {Phys.
  Rev. B}\ }\textbf {\bibinfo {volume} {76}},\ \bibinfo {pages} {094522}
  (\bibinfo {year} {2007})}\BibitemShut {NoStop}%
\bibitem [{\citenamefont {Bannykh}\ \emph {et~al.}(2009)\citenamefont
  {Bannykh}, \citenamefont {Pfeiffer}, \citenamefont {Stolyarov}, \citenamefont
  {Batov}, \citenamefont {Ryazanov},\ and\ \citenamefont
  {Weides}}]{Bannykh2009}%
  \BibitemOpen
  \bibfield  {author} {\bibinfo {author} {\bibfnamefont {A.~A.}\ \bibnamefont
  {Bannykh}}, \bibinfo {author} {\bibfnamefont {J.}~\bibnamefont {Pfeiffer}},
  \bibinfo {author} {\bibfnamefont {V.~S.}\ \bibnamefont {Stolyarov}}, \bibinfo
  {author} {\bibfnamefont {I.~E.}\ \bibnamefont {Batov}}, \bibinfo {author}
  {\bibfnamefont {V.~V.}\ \bibnamefont {Ryazanov}}, \ and\ \bibinfo {author}
  {\bibfnamefont {M.}~\bibnamefont {Weides}},\ }\href {\doibase
  10.1103/PhysRevB.79.054501} {\bibfield  {journal} {\bibinfo  {journal} {Phys.
  Rev. B}\ }\textbf {\bibinfo {volume} {79}},\ \bibinfo {pages} {054501}
  (\bibinfo {year} {2009})}\BibitemShut {NoStop}%
\bibitem [{\citenamefont {Khaire}\ \emph {et~al.}(2009)\citenamefont {Khaire},
  \citenamefont {Pratt},\ and\ \citenamefont {Birge}}]{Khaire2009}%
  \BibitemOpen
  \bibfield  {author} {\bibinfo {author} {\bibfnamefont {T.~S.}\ \bibnamefont
  {Khaire}}, \bibinfo {author} {\bibfnamefont {W.~P.}\ \bibnamefont {Pratt}}, \
  and\ \bibinfo {author} {\bibfnamefont {N.~O.}\ \bibnamefont {Birge}},\ }\href
  {\doibase 10.1103/PhysRevB.79.094523} {\bibfield  {journal} {\bibinfo
  {journal} {Phys. Rev. B}\ }\textbf {\bibinfo {volume} {79}},\ \bibinfo
  {pages} {094523} (\bibinfo {year} {2009})}\BibitemShut {NoStop}%
\bibitem [{\citenamefont {Niedzielski}\ \emph {et~al.}(2015)\citenamefont
  {Niedzielski}, \citenamefont {Gingrich}, \citenamefont {Loloee},
  \citenamefont {Pratt},\ and\ \citenamefont {Birge}}]{Niedzielski2015}%
  \BibitemOpen
  \bibfield  {author} {\bibinfo {author} {\bibfnamefont {B.~M.}\ \bibnamefont
  {Niedzielski}}, \bibinfo {author} {\bibfnamefont {E.~C.}\ \bibnamefont
  {Gingrich}}, \bibinfo {author} {\bibfnamefont {R.}~\bibnamefont {Loloee}},
  \bibinfo {author} {\bibfnamefont {W.~P.}\ \bibnamefont {Pratt}}, \ and\
  \bibinfo {author} {\bibfnamefont {N.~O.}\ \bibnamefont {Birge}},\ }\href@noop
  {} {\bibfield  {journal} {\bibinfo  {journal} {Supercond. Sci. Technol.}\
  }\textbf {\bibinfo {volume} {28}},\ \bibinfo {pages} {085012} (\bibinfo
  {year} {2015})}\BibitemShut {NoStop}%
\bibitem [{\citenamefont {Glick}\ \emph {et~al.}(2017)\citenamefont {Glick},
  \citenamefont {Khasawneh}, \citenamefont {Niedzielski}, \citenamefont
  {Loloee}, \citenamefont {Pratt}, \citenamefont {Birge}, \citenamefont
  {Gingrich}, \citenamefont {Kotula},\ and\ \citenamefont
  {Missert}}]{Glick2016}%
  \BibitemOpen
  \bibfield  {author} {\bibinfo {author} {\bibfnamefont {J.~A.}\ \bibnamefont
  {Glick}}, \bibinfo {author} {\bibfnamefont {M.~A.}\ \bibnamefont
  {Khasawneh}}, \bibinfo {author} {\bibfnamefont {B.~M.}\ \bibnamefont
  {Niedzielski}}, \bibinfo {author} {\bibfnamefont {R.}~\bibnamefont {Loloee}},
  \bibinfo {author} {\bibfnamefont {W.~P.}\ \bibnamefont {Pratt}}, \bibinfo
  {author} {\bibfnamefont {N.~O.}\ \bibnamefont {Birge}}, \bibinfo {author}
  {\bibfnamefont {E.~C.}\ \bibnamefont {Gingrich}}, \bibinfo {author}
  {\bibfnamefont {P.~G.}\ \bibnamefont {Kotula}}, \ and\ \bibinfo {author}
  {\bibfnamefont {N.}~\bibnamefont {Missert}},\ }\href {\doibase
  10.1063/1.4989392} {\bibfield  {journal} {\bibinfo  {journal} {J. App.
  Phys.}\ }\textbf {\bibinfo {volume} {122}},\ \bibinfo {pages} {133906}
  (\bibinfo {year} {2017})}\BibitemShut {NoStop}%
\bibitem [{\citenamefont {Eschrig}\ \emph {et~al.}(2003)\citenamefont
  {Eschrig}, \citenamefont {Kopu}, \citenamefont {Cuevas},\ and\ \citenamefont
  {Sch\"on}}]{Eschrig2003}%
  \BibitemOpen
  \bibfield  {author} {\bibinfo {author} {\bibfnamefont {M.}~\bibnamefont
  {Eschrig}}, \bibinfo {author} {\bibfnamefont {J.}~\bibnamefont {Kopu}},
  \bibinfo {author} {\bibfnamefont {J.~C.}\ \bibnamefont {Cuevas}}, \ and\
  \bibinfo {author} {\bibfnamefont {G.}~\bibnamefont {Sch\"on}},\ }\href
  {\doibase 10.1103/PhysRevLett.90.137003} {\bibfield  {journal} {\bibinfo
  {journal} {Phys. Rev. Lett.}\ }\textbf {\bibinfo {volume} {90}},\ \bibinfo
  {pages} {137003} (\bibinfo {year} {2003})}\BibitemShut {NoStop}%
\bibitem [{\citenamefont {Bergeret}\ \emph {et~al.}(2001)\citenamefont
  {Bergeret}, \citenamefont {Volkov},\ and\ \citenamefont
  {Efetov}}]{Bergeret2001}%
  \BibitemOpen
  \bibfield  {author} {\bibinfo {author} {\bibfnamefont {F.~S.}\ \bibnamefont
  {Bergeret}}, \bibinfo {author} {\bibfnamefont {A.~F.}\ \bibnamefont
  {Volkov}}, \ and\ \bibinfo {author} {\bibfnamefont {K.~B.}\ \bibnamefont
  {Efetov}},\ }\href {\doibase 10.1103/PhysRevB.64.134506} {\bibfield
  {journal} {\bibinfo  {journal} {Phys. Rev. B}\ }\textbf {\bibinfo {volume}
  {64}},\ \bibinfo {pages} {134506} (\bibinfo {year} {2001})}\BibitemShut
  {NoStop}%
\bibitem [{\citenamefont {Kadigrobov}\ \emph {et~al.}(2001)\citenamefont
  {Kadigrobov}, \citenamefont {Shekhter},\ and\ \citenamefont
  {Jonson}}]{Kadigrobov2001}%
  \BibitemOpen
  \bibfield  {author} {\bibinfo {author} {\bibfnamefont {A.}~\bibnamefont
  {Kadigrobov}}, \bibinfo {author} {\bibfnamefont {R.~I.}\ \bibnamefont
  {Shekhter}}, \ and\ \bibinfo {author} {\bibfnamefont {M.}~\bibnamefont
  {Jonson}},\ }\href {http://stacks.iop.org/0295-5075/54/i=3/a=394} {\bibfield
  {journal} {\bibinfo  {journal} {EPL (Europhysics Letters)}\ }\textbf
  {\bibinfo {volume} {54}},\ \bibinfo {pages} {394} (\bibinfo {year}
  {2001})}\BibitemShut {NoStop}%
\bibitem [{\citenamefont {Volkov}\ \emph {et~al.}(2003)\citenamefont {Volkov},
  \citenamefont {Bergeret},\ and\ \citenamefont {Efetov}}]{Volkov2003}%
  \BibitemOpen
  \bibfield  {author} {\bibinfo {author} {\bibfnamefont {A.~F.}\ \bibnamefont
  {Volkov}}, \bibinfo {author} {\bibfnamefont {F.~S.}\ \bibnamefont
  {Bergeret}}, \ and\ \bibinfo {author} {\bibfnamefont {K.~B.}\ \bibnamefont
  {Efetov}},\ }\href {\doibase 10.1103/PhysRevLett.90.117006} {\bibfield
  {journal} {\bibinfo  {journal} {Phys. Rev. Lett.}\ }\textbf {\bibinfo
  {volume} {90}},\ \bibinfo {pages} {117006} (\bibinfo {year}
  {2003})}\BibitemShut {NoStop}%
\bibitem [{\citenamefont {Bergeret}\ \emph {et~al.}(2005)\citenamefont
  {Bergeret}, \citenamefont {Volkov},\ and\ \citenamefont
  {Efetov}}]{Bergeret2005}%
  \BibitemOpen
  \bibfield  {author} {\bibinfo {author} {\bibfnamefont {F.~S.}\ \bibnamefont
  {Bergeret}}, \bibinfo {author} {\bibfnamefont {A.~F.}\ \bibnamefont
  {Volkov}}, \ and\ \bibinfo {author} {\bibfnamefont {K.~B.}\ \bibnamefont
  {Efetov}},\ }\href {\doibase 10.1103/RevModPhys.77.1321} {\bibfield
  {journal} {\bibinfo  {journal} {Rev. Mod. Phys.}\ }\textbf {\bibinfo {volume}
  {77}},\ \bibinfo {pages} {1321} (\bibinfo {year} {2005})}\BibitemShut
  {NoStop}%
\bibitem [{\citenamefont {Keizer}\ \emph {et~al.}(2006)\citenamefont {Keizer},
  \citenamefont {Goennenwein}, \citenamefont {Klapwijk}, \citenamefont {Miao},
  \citenamefont {Xiao},\ and\ \citenamefont {Gupta}}]{Keizer2006}%
  \BibitemOpen
  \bibfield  {author} {\bibinfo {author} {\bibfnamefont {R.~S.}\ \bibnamefont
  {Keizer}}, \bibinfo {author} {\bibfnamefont {S.~T.~B.}\ \bibnamefont
  {Goennenwein}}, \bibinfo {author} {\bibfnamefont {T.~M.}\ \bibnamefont
  {Klapwijk}}, \bibinfo {author} {\bibfnamefont {G.}~\bibnamefont {Miao}},
  \bibinfo {author} {\bibfnamefont {G.}~\bibnamefont {Xiao}}, \ and\ \bibinfo
  {author} {\bibfnamefont {A.}~\bibnamefont {Gupta}},\ }\href {\doibase
  10.1038/nature04499} {\bibfield  {journal} {\bibinfo  {journal} {Nature}\
  }\textbf {\bibinfo {volume} {439}},\ \bibinfo {pages} {825} (\bibinfo {year}
  {2006})}\BibitemShut {NoStop}%
\bibitem [{\citenamefont {Sosnin}\ \emph {et~al.}(2006)\citenamefont {Sosnin},
  \citenamefont {Cho}, \citenamefont {Petrashov},\ and\ \citenamefont
  {Volkov}}]{Sosnin2006}%
  \BibitemOpen
  \bibfield  {author} {\bibinfo {author} {\bibfnamefont {I.}~\bibnamefont
  {Sosnin}}, \bibinfo {author} {\bibfnamefont {H.}~\bibnamefont {Cho}},
  \bibinfo {author} {\bibfnamefont {V.~T.}\ \bibnamefont {Petrashov}}, \ and\
  \bibinfo {author} {\bibfnamefont {A.~F.}\ \bibnamefont {Volkov}},\ }\href
  {\doibase 10.1103/PhysRevLett.96.157002} {\bibfield  {journal} {\bibinfo
  {journal} {Phys. Rev. Lett.}\ }\textbf {\bibinfo {volume} {96}},\ \bibinfo
  {pages} {157002} (\bibinfo {year} {2006})}\BibitemShut {NoStop}%
\bibitem [{\citenamefont {Khaire}\ \emph {et~al.}(2010)\citenamefont {Khaire},
  \citenamefont {Khasawneh}, \citenamefont {Pratt},\ and\ \citenamefont
  {Birge}}]{Khaire2010}%
  \BibitemOpen
  \bibfield  {author} {\bibinfo {author} {\bibfnamefont {T.~S.}\ \bibnamefont
  {Khaire}}, \bibinfo {author} {\bibfnamefont {M.~A.}\ \bibnamefont
  {Khasawneh}}, \bibinfo {author} {\bibfnamefont {W.~P.}\ \bibnamefont
  {Pratt}}, \ and\ \bibinfo {author} {\bibfnamefont {N.~O.}\ \bibnamefont
  {Birge}},\ }\href {\doibase 10.1103/PhysRevLett.104.137002} {\bibfield
  {journal} {\bibinfo  {journal} {Phys. Rev. Lett.}\ }\textbf {\bibinfo
  {volume} {104}},\ \bibinfo {pages} {137002} (\bibinfo {year}
  {2010})}\BibitemShut {NoStop}%
\bibitem [{\citenamefont {Robinson}\ \emph {et~al.}(2010)\citenamefont
  {Robinson}, \citenamefont {Witt},\ and\ \citenamefont
  {Blamire}}]{Robinson2010Science}%
  \BibitemOpen
  \bibfield  {author} {\bibinfo {author} {\bibfnamefont {J.~W.~A.}\
  \bibnamefont {Robinson}}, \bibinfo {author} {\bibfnamefont {J.~D.~S.}\
  \bibnamefont {Witt}}, \ and\ \bibinfo {author} {\bibfnamefont {M.~G.}\
  \bibnamefont {Blamire}},\ }\href {\doibase 10.1126/science.1189246}
  {\bibfield  {journal} {\bibinfo  {journal} {Science}\ }\textbf {\bibinfo
  {volume} {329}},\ \bibinfo {pages} {59} (\bibinfo {year} {2010})}\BibitemShut
  {NoStop}%
\bibitem [{\citenamefont {Sprungmann}\ \emph {et~al.}(2010)\citenamefont
  {Sprungmann}, \citenamefont {Westerholt}, \citenamefont {Zabel},
  \citenamefont {Weides},\ and\ \citenamefont {Kohlstedt}}]{Sprungmann2010}%
  \BibitemOpen
  \bibfield  {author} {\bibinfo {author} {\bibfnamefont {D.}~\bibnamefont
  {Sprungmann}}, \bibinfo {author} {\bibfnamefont {K.}~\bibnamefont
  {Westerholt}}, \bibinfo {author} {\bibfnamefont {H.}~\bibnamefont {Zabel}},
  \bibinfo {author} {\bibfnamefont {M.}~\bibnamefont {Weides}}, \ and\ \bibinfo
  {author} {\bibfnamefont {H.}~\bibnamefont {Kohlstedt}},\ }\href {\doibase
  10.1103/PhysRevB.82.060505} {\bibfield  {journal} {\bibinfo  {journal} {Phys.
  Rev. B}\ }\textbf {\bibinfo {volume} {82}},\ \bibinfo {pages} {060505}
  (\bibinfo {year} {2010})}\BibitemShut {NoStop}%
\bibitem [{\citenamefont {Anwar}\ \emph {et~al.}(2010)\citenamefont {Anwar},
  \citenamefont {Czeschka}, \citenamefont {Hesselberth}, \citenamefont
  {Porcu},\ and\ \citenamefont {Aarts}}]{Anwar2010}%
  \BibitemOpen
  \bibfield  {author} {\bibinfo {author} {\bibfnamefont {M.~S.}\ \bibnamefont
  {Anwar}}, \bibinfo {author} {\bibfnamefont {F.}~\bibnamefont {Czeschka}},
  \bibinfo {author} {\bibfnamefont {M.}~\bibnamefont {Hesselberth}}, \bibinfo
  {author} {\bibfnamefont {M.}~\bibnamefont {Porcu}}, \ and\ \bibinfo {author}
  {\bibfnamefont {J.}~\bibnamefont {Aarts}},\ }\href {\doibase
  10.1103/PhysRevB.82.100501} {\bibfield  {journal} {\bibinfo  {journal} {Phys.
  Rev. B}\ }\textbf {\bibinfo {volume} {82}},\ \bibinfo {pages} {100501}
  (\bibinfo {year} {2010})}\BibitemShut {NoStop}%
\bibitem [{\citenamefont {Wang}\ \emph {et~al.}(2010)\citenamefont {Wang},
  \citenamefont {Singh}, \citenamefont {Tian}, \citenamefont {Kumar},
  \citenamefont {Liu}, \citenamefont {Shi}, \citenamefont {Jain}, \citenamefont
  {Samarth}, \citenamefont {Mallouk},\ and\ \citenamefont {Chan}}]{Wang2010}%
  \BibitemOpen
  \bibfield  {author} {\bibinfo {author} {\bibfnamefont {J.}~\bibnamefont
  {Wang}}, \bibinfo {author} {\bibfnamefont {M.}~\bibnamefont {Singh}},
  \bibinfo {author} {\bibfnamefont {M.}~\bibnamefont {Tian}}, \bibinfo {author}
  {\bibfnamefont {N.}~\bibnamefont {Kumar}}, \bibinfo {author} {\bibfnamefont
  {B.}~\bibnamefont {Liu}}, \bibinfo {author} {\bibfnamefont {C.}~\bibnamefont
  {Shi}}, \bibinfo {author} {\bibfnamefont {J.~K.}\ \bibnamefont {Jain}},
  \bibinfo {author} {\bibfnamefont {N.}~\bibnamefont {Samarth}}, \bibinfo
  {author} {\bibfnamefont {T.~E.}\ \bibnamefont {Mallouk}}, \ and\ \bibinfo
  {author} {\bibfnamefont {M.~H.~W.}\ \bibnamefont {Chan}},\ }\href {\doibase
  10.1038/nphys1621} {\bibfield  {journal} {\bibinfo  {journal} {Nat. Phys.}\
  }\textbf {\bibinfo {volume} {6}},\ \bibinfo {pages} {389} (\bibinfo {year}
  {2010})}\BibitemShut {NoStop}%
\bibitem [{\citenamefont {Klose}\ \emph {et~al.}(2012)\citenamefont {Klose},
  \citenamefont {Khaire}, \citenamefont {Wang}, \citenamefont {Pratt},
  \citenamefont {Birge}, \citenamefont {McMorran}, \citenamefont {Ginley},
  \citenamefont {Borchers}, \citenamefont {Kirby}, \citenamefont {Maranville},\
  and\ \citenamefont {Unguris}}]{Klose2012}%
  \BibitemOpen
  \bibfield  {author} {\bibinfo {author} {\bibfnamefont {C.}~\bibnamefont
  {Klose}}, \bibinfo {author} {\bibfnamefont {T.~S.}\ \bibnamefont {Khaire}},
  \bibinfo {author} {\bibfnamefont {Y.}~\bibnamefont {Wang}}, \bibinfo {author}
  {\bibfnamefont {W.~P.}\ \bibnamefont {Pratt}}, \bibinfo {author}
  {\bibfnamefont {N.~O.}\ \bibnamefont {Birge}}, \bibinfo {author}
  {\bibfnamefont {B.~J.}\ \bibnamefont {McMorran}}, \bibinfo {author}
  {\bibfnamefont {T.~P.}\ \bibnamefont {Ginley}}, \bibinfo {author}
  {\bibfnamefont {J.~A.}\ \bibnamefont {Borchers}}, \bibinfo {author}
  {\bibfnamefont {B.~J.}\ \bibnamefont {Kirby}}, \bibinfo {author}
  {\bibfnamefont {B.~B.}\ \bibnamefont {Maranville}}, \ and\ \bibinfo {author}
  {\bibfnamefont {J.}~\bibnamefont {Unguris}},\ }\href {\doibase
  10.1103/PhysRevLett.108.127002} {\bibfield  {journal} {\bibinfo  {journal}
  {Phys. Rev. Lett.}\ }\textbf {\bibinfo {volume} {108}},\ \bibinfo {pages}
  {127002} (\bibinfo {year} {2012})}\BibitemShut {NoStop}%
\bibitem [{\citenamefont {Leksin}\ \emph {et~al.}(2012)\citenamefont {Leksin},
  \citenamefont {Garif'yanov}, \citenamefont {Garifullin}, \citenamefont
  {Fominov}, \citenamefont {Schumann}, \citenamefont {Krupskaya}, \citenamefont
  {Kataev}, \citenamefont {Schmidt},\ and\ \citenamefont
  {B\"uchner}}]{Leksin2012}%
  \BibitemOpen
  \bibfield  {author} {\bibinfo {author} {\bibfnamefont {P.~V.}\ \bibnamefont
  {Leksin}}, \bibinfo {author} {\bibfnamefont {N.~N.}\ \bibnamefont
  {Garif'yanov}}, \bibinfo {author} {\bibfnamefont {I.~A.}\ \bibnamefont
  {Garifullin}}, \bibinfo {author} {\bibfnamefont {Y.~V.}\ \bibnamefont
  {Fominov}}, \bibinfo {author} {\bibfnamefont {J.}~\bibnamefont {Schumann}},
  \bibinfo {author} {\bibfnamefont {Y.}~\bibnamefont {Krupskaya}}, \bibinfo
  {author} {\bibfnamefont {V.}~\bibnamefont {Kataev}}, \bibinfo {author}
  {\bibfnamefont {O.~G.}\ \bibnamefont {Schmidt}}, \ and\ \bibinfo {author}
  {\bibfnamefont {B.}~\bibnamefont {B\"uchner}},\ }\href {\doibase
  10.1103/PhysRevLett.109.057005} {\bibfield  {journal} {\bibinfo  {journal}
  {Phys. Rev. Lett.}\ }\textbf {\bibinfo {volume} {109}},\ \bibinfo {pages}
  {057005} (\bibinfo {year} {2012})}\BibitemShut {NoStop}%
\bibitem [{\citenamefont {Zdravkov}\ \emph {et~al.}(2013)\citenamefont
  {Zdravkov}, \citenamefont {Kehrle}, \citenamefont {Obermeier}, \citenamefont
  {Lenk}, \citenamefont {Krug~von Nidda}, \citenamefont {M\"uller},
  \citenamefont {Kupriyanov}, \citenamefont {Sidorenko}, \citenamefont {Horn},
  \citenamefont {Tidecks},\ and\ \citenamefont {Tagirov}}]{Zdravkov2013}%
  \BibitemOpen
  \bibfield  {author} {\bibinfo {author} {\bibfnamefont {V.~I.}\ \bibnamefont
  {Zdravkov}}, \bibinfo {author} {\bibfnamefont {J.}~\bibnamefont {Kehrle}},
  \bibinfo {author} {\bibfnamefont {G.}~\bibnamefont {Obermeier}}, \bibinfo
  {author} {\bibfnamefont {D.}~\bibnamefont {Lenk}}, \bibinfo {author}
  {\bibfnamefont {H.-A.}\ \bibnamefont {Krug~von Nidda}}, \bibinfo {author}
  {\bibfnamefont {C.}~\bibnamefont {M\"uller}}, \bibinfo {author}
  {\bibfnamefont {M.~Y.}\ \bibnamefont {Kupriyanov}}, \bibinfo {author}
  {\bibfnamefont {A.~S.}\ \bibnamefont {Sidorenko}}, \bibinfo {author}
  {\bibfnamefont {S.}~\bibnamefont {Horn}}, \bibinfo {author} {\bibfnamefont
  {R.}~\bibnamefont {Tidecks}}, \ and\ \bibinfo {author} {\bibfnamefont
  {L.~R.}\ \bibnamefont {Tagirov}},\ }\href {\doibase
  10.1103/PhysRevB.87.144507} {\bibfield  {journal} {\bibinfo  {journal} {Phys.
  Rev. B}\ }\textbf {\bibinfo {volume} {87}},\ \bibinfo {pages} {144507}
  (\bibinfo {year} {2013})}\BibitemShut {NoStop}%
\bibitem [{\citenamefont {Banerjee}\ \emph
  {et~al.}(2014{\natexlab{a}})\citenamefont {Banerjee}, \citenamefont {Smiet},
  \citenamefont {Smits}, \citenamefont {Ozaeta}, \citenamefont {Bergeret},
  \citenamefont {Blamire},\ and\ \citenamefont
  {Robinson}}]{Banerjee2014-TripletSpinValve}%
  \BibitemOpen
  \bibfield  {author} {\bibinfo {author} {\bibfnamefont {N.}~\bibnamefont
  {Banerjee}}, \bibinfo {author} {\bibfnamefont {C.~B.}\ \bibnamefont {Smiet}},
  \bibinfo {author} {\bibfnamefont {R.~G.~J.}\ \bibnamefont {Smits}}, \bibinfo
  {author} {\bibfnamefont {A.}~\bibnamefont {Ozaeta}}, \bibinfo {author}
  {\bibfnamefont {F.~S.}\ \bibnamefont {Bergeret}}, \bibinfo {author}
  {\bibfnamefont {M.~G.}\ \bibnamefont {Blamire}}, \ and\ \bibinfo {author}
  {\bibfnamefont {J.~W.~A.}\ \bibnamefont {Robinson}},\ }\href {\doibase
  10.1038/ncomms4048} {\bibfield  {journal} {\bibinfo  {journal} {Nat. Comm.}\
  }\textbf {\bibinfo {volume} {5}},\ \bibinfo {pages} {3048} (\bibinfo {year}
  {2014}{\natexlab{a}})}\BibitemShut {NoStop}%
\bibitem [{\citenamefont {Banerjee}\ \emph
  {et~al.}(2014{\natexlab{b}})\citenamefont {Banerjee}, \citenamefont
  {Robinson},\ and\ \citenamefont {Blamire}}]{Banerjee2014}%
  \BibitemOpen
  \bibfield  {author} {\bibinfo {author} {\bibfnamefont {N.}~\bibnamefont
  {Banerjee}}, \bibinfo {author} {\bibfnamefont {J.~W.~A.}\ \bibnamefont
  {Robinson}}, \ and\ \bibinfo {author} {\bibfnamefont {M.~G.}\ \bibnamefont
  {Blamire}},\ }\href {http://dx.doi.org/10.1038/ncomms5771} {\bibfield
  {journal} {\bibinfo  {journal} {Nat. Comm.}\ }\textbf {\bibinfo {volume}
  {5}},\ \bibinfo {pages} {4771 EP} (\bibinfo {year}
  {2014}{\natexlab{b}})}\BibitemShut {NoStop}%
\bibitem [{\citenamefont {Iovan}\ \emph {et~al.}(2014)\citenamefont {Iovan},
  \citenamefont {Golod},\ and\ \citenamefont {Krasnov}}]{Iovan2014}%
  \BibitemOpen
  \bibfield  {author} {\bibinfo {author} {\bibfnamefont {A.}~\bibnamefont
  {Iovan}}, \bibinfo {author} {\bibfnamefont {T.}~\bibnamefont {Golod}}, \ and\
  \bibinfo {author} {\bibfnamefont {V.~M.}\ \bibnamefont {Krasnov}},\ }\href
  {\doibase 10.1103/PhysRevB.90.134514} {\bibfield  {journal} {\bibinfo
  {journal} {Phys. Rev. B}\ }\textbf {\bibinfo {volume} {90}},\ \bibinfo
  {pages} {134514} (\bibinfo {year} {2014})}\BibitemShut {NoStop}%
\bibitem [{\citenamefont {Wang}\ \emph {et~al.}(2014)\citenamefont {Wang},
  \citenamefont {Di~Bernardo}, \citenamefont {Banerjee}, \citenamefont {Wells},
  \citenamefont {Bergeret}, \citenamefont {Blamire},\ and\ \citenamefont
  {Robinson}}]{Wang2014}%
  \BibitemOpen
  \bibfield  {author} {\bibinfo {author} {\bibfnamefont {X.~L.}\ \bibnamefont
  {Wang}}, \bibinfo {author} {\bibfnamefont {A.}~\bibnamefont {Di~Bernardo}},
  \bibinfo {author} {\bibfnamefont {N.}~\bibnamefont {Banerjee}}, \bibinfo
  {author} {\bibfnamefont {A.}~\bibnamefont {Wells}}, \bibinfo {author}
  {\bibfnamefont {F.~S.}\ \bibnamefont {Bergeret}}, \bibinfo {author}
  {\bibfnamefont {M.~G.}\ \bibnamefont {Blamire}}, \ and\ \bibinfo {author}
  {\bibfnamefont {J.~W.~A.}\ \bibnamefont {Robinson}},\ }\href {\doibase
  10.1103/PhysRevB.89.140508} {\bibfield  {journal} {\bibinfo  {journal} {Phys.
  Rev. B}\ }\textbf {\bibinfo {volume} {89}},\ \bibinfo {pages} {140508}
  (\bibinfo {year} {2014})}\BibitemShut {NoStop}%
\bibitem [{\citenamefont {Jara}\ \emph {et~al.}(2014)\citenamefont {Jara},
  \citenamefont {Safranski}, \citenamefont {Krivorotov}, \citenamefont {Wu},
  \citenamefont {Malmi-Kakkada}, \citenamefont {Valls},\ and\ \citenamefont
  {Halterman}}]{Jara2014}%
  \BibitemOpen
  \bibfield  {author} {\bibinfo {author} {\bibfnamefont {A.~A.}\ \bibnamefont
  {Jara}}, \bibinfo {author} {\bibfnamefont {C.}~\bibnamefont {Safranski}},
  \bibinfo {author} {\bibfnamefont {I.~N.}\ \bibnamefont {Krivorotov}},
  \bibinfo {author} {\bibfnamefont {C.}~\bibnamefont {Wu}}, \bibinfo {author}
  {\bibfnamefont {A.~N.}\ \bibnamefont {Malmi-Kakkada}}, \bibinfo {author}
  {\bibfnamefont {O.~T.}\ \bibnamefont {Valls}}, \ and\ \bibinfo {author}
  {\bibfnamefont {K.}~\bibnamefont {Halterman}},\ }\href {\doibase
  10.1103/PhysRevB.89.184502} {\bibfield  {journal} {\bibinfo  {journal} {Phys.
  Rev. B}\ }\textbf {\bibinfo {volume} {89}},\ \bibinfo {pages} {184502}
  (\bibinfo {year} {2014})}\BibitemShut {NoStop}%
\bibitem [{\citenamefont {Flokstra}\ \emph {et~al.}(2015)\citenamefont
  {Flokstra}, \citenamefont {Cunningham}, \citenamefont {Kim}, \citenamefont
  {Satchell}, \citenamefont {Burnell}, \citenamefont {Curran}, \citenamefont
  {Bending}, \citenamefont {Kinane}, \citenamefont {Cooper}, \citenamefont
  {Langridge}, \citenamefont {Isidori}, \citenamefont {Pugach}, \citenamefont
  {Eschrig},\ and\ \citenamefont {Lee}}]{Flokstra2015}%
  \BibitemOpen
  \bibfield  {author} {\bibinfo {author} {\bibfnamefont {M.~G.}\ \bibnamefont
  {Flokstra}}, \bibinfo {author} {\bibfnamefont {T.~C.}\ \bibnamefont
  {Cunningham}}, \bibinfo {author} {\bibfnamefont {J.}~\bibnamefont {Kim}},
  \bibinfo {author} {\bibfnamefont {N.}~\bibnamefont {Satchell}}, \bibinfo
  {author} {\bibfnamefont {G.}~\bibnamefont {Burnell}}, \bibinfo {author}
  {\bibfnamefont {P.~J.}\ \bibnamefont {Curran}}, \bibinfo {author}
  {\bibfnamefont {S.~J.}\ \bibnamefont {Bending}}, \bibinfo {author}
  {\bibfnamefont {C.~J.}\ \bibnamefont {Kinane}}, \bibinfo {author}
  {\bibfnamefont {J.~F.~K.}\ \bibnamefont {Cooper}}, \bibinfo {author}
  {\bibfnamefont {S.}~\bibnamefont {Langridge}}, \bibinfo {author}
  {\bibfnamefont {A.}~\bibnamefont {Isidori}}, \bibinfo {author} {\bibfnamefont
  {N.}~\bibnamefont {Pugach}}, \bibinfo {author} {\bibfnamefont
  {M.}~\bibnamefont {Eschrig}}, \ and\ \bibinfo {author} {\bibfnamefont
  {S.~L.}\ \bibnamefont {Lee}},\ }\href {\doibase 10.1103/PhysRevB.91.060501}
  {\bibfield  {journal} {\bibinfo  {journal} {Phys. Rev. B}\ }\textbf {\bibinfo
  {volume} {91}},\ \bibinfo {pages} {060501} (\bibinfo {year}
  {2015})}\BibitemShut {NoStop}%
\bibitem [{\citenamefont {Singh}\ \emph {et~al.}(2015)\citenamefont {Singh},
  \citenamefont {Voltan}, \citenamefont {Lahabi},\ and\ \citenamefont
  {Aarts}}]{Singh2015}%
  \BibitemOpen
  \bibfield  {author} {\bibinfo {author} {\bibfnamefont {A.}~\bibnamefont
  {Singh}}, \bibinfo {author} {\bibfnamefont {S.}~\bibnamefont {Voltan}},
  \bibinfo {author} {\bibfnamefont {K.}~\bibnamefont {Lahabi}}, \ and\ \bibinfo
  {author} {\bibfnamefont {J.}~\bibnamefont {Aarts}},\ }\href {\doibase
  10.1103/PhysRevX.5.021019} {\bibfield  {journal} {\bibinfo  {journal} {Phys.
  Rev. X}\ }\textbf {\bibinfo {volume} {5}},\ \bibinfo {pages} {021019}
  (\bibinfo {year} {2015})}\BibitemShut {NoStop}%
\bibitem [{\citenamefont {Linder}\ and\ \citenamefont
  {Robinson}(2015)}]{Linder2015}%
  \BibitemOpen
  \bibfield  {author} {\bibinfo {author} {\bibfnamefont {J.}~\bibnamefont
  {Linder}}\ and\ \bibinfo {author} {\bibfnamefont {J.~W.~A.}\ \bibnamefont
  {Robinson}},\ }\href {http://dx.doi.org/10.1038/nphys3242} {\bibfield
  {journal} {\bibinfo  {journal} {Nat. Phys.}\ }\textbf {\bibinfo {volume}
  {11}},\ \bibinfo {pages} {307} (\bibinfo {year} {2015})},\ \bibinfo {note}
  {review}\BibitemShut {NoStop}%
\bibitem [{\citenamefont {Eschrig}(2015)}]{Eschrig2015}%
  \BibitemOpen
  \bibfield  {author} {\bibinfo {author} {\bibfnamefont {M.}~\bibnamefont
  {Eschrig}},\ }\href {http://stacks.iop.org/0034-4885/78/i=10/a=104501}
  {\bibfield  {journal} {\bibinfo  {journal} {Reports on Progress in Physics}\
  }\textbf {\bibinfo {volume} {78}},\ \bibinfo {pages} {104501} (\bibinfo
  {year} {2015})}\BibitemShut {NoStop}%
\bibitem [{\citenamefont {Di~Bernardo}\ \emph {et~al.}(2015)\citenamefont
  {Di~Bernardo}, \citenamefont {Diesch}, \citenamefont {Gu}, \citenamefont
  {Linder}, \citenamefont {Divitini}, \citenamefont {Ducati}, \citenamefont
  {Scheer}, \citenamefont {Blamire},\ and\ \citenamefont
  {Robinson}}]{DiBernardo2015}%
  \BibitemOpen
  \bibfield  {author} {\bibinfo {author} {\bibfnamefont {A.}~\bibnamefont
  {Di~Bernardo}}, \bibinfo {author} {\bibfnamefont {S.}~\bibnamefont {Diesch}},
  \bibinfo {author} {\bibfnamefont {Y.}~\bibnamefont {Gu}}, \bibinfo {author}
  {\bibfnamefont {J.}~\bibnamefont {Linder}}, \bibinfo {author} {\bibfnamefont
  {G.}~\bibnamefont {Divitini}}, \bibinfo {author} {\bibfnamefont
  {C.}~\bibnamefont {Ducati}}, \bibinfo {author} {\bibfnamefont
  {E.}~\bibnamefont {Scheer}}, \bibinfo {author} {\bibfnamefont {M.~G.}\
  \bibnamefont {Blamire}}, \ and\ \bibinfo {author} {\bibfnamefont {J.~W.~A.}\
  \bibnamefont {Robinson}},\ }\href {http://dx.doi.org/10.1038/ncomms9053}
  {\bibfield  {journal} {\bibinfo  {journal} {Nat. Comm.}\ }\textbf {\bibinfo
  {volume} {6}},\ \bibinfo {pages} {8053} (\bibinfo {year} {2015})}\BibitemShut
  {NoStop}%
\bibitem [{\citenamefont {Feng}\ \emph {et~al.}(2017)\citenamefont {Feng},
  \citenamefont {Robinson},\ and\ \citenamefont {Blamire}}]{Feng2017}%
  \BibitemOpen
  \bibfield  {author} {\bibinfo {author} {\bibfnamefont {Z.}~\bibnamefont
  {Feng}}, \bibinfo {author} {\bibfnamefont {J.~W.~A.}\ \bibnamefont
  {Robinson}}, \ and\ \bibinfo {author} {\bibfnamefont {M.~G.}\ \bibnamefont
  {Blamire}},\ }\href {\doibase 10.1063/1.4995434} {\bibfield  {journal}
  {\bibinfo  {journal} {Appl. Phys. Lett.}\ }\textbf {\bibinfo {volume}
  {111}},\ \bibinfo {pages} {042602} (\bibinfo {year} {2017})}\BibitemShut
  {NoStop}%
\bibitem [{\citenamefont {Cirillo}\ \emph {et~al.}(2017)\citenamefont
  {Cirillo}, \citenamefont {Voltan}, \citenamefont {Ilyina}, \citenamefont
  {Hernandez}, \citenamefont {Garcia-Santiago}, \citenamefont {Aarts},\ and\
  \citenamefont {Attanasio}}]{Cirillo2017}%
  \BibitemOpen
  \bibfield  {author} {\bibinfo {author} {\bibfnamefont {C.}~\bibnamefont
  {Cirillo}}, \bibinfo {author} {\bibfnamefont {S.}~\bibnamefont {Voltan}},
  \bibinfo {author} {\bibfnamefont {E.~A.}\ \bibnamefont {Ilyina}}, \bibinfo
  {author} {\bibfnamefont {J.~M.}\ \bibnamefont {Hernandez}}, \bibinfo {author}
  {\bibfnamefont {A.}~\bibnamefont {Garcia-Santiago}}, \bibinfo {author}
  {\bibfnamefont {J.}~\bibnamefont {Aarts}}, \ and\ \bibinfo {author}
  {\bibfnamefont {C.}~\bibnamefont {Attanasio}},\ }\href
  {http://stacks.iop.org/1367-2630/19/i=2/a=023037} {\bibfield  {journal}
  {\bibinfo  {journal} {New J. Phys.}\ }\textbf {\bibinfo {volume} {19}},\
  \bibinfo {pages} {023037} (\bibinfo {year} {2017})}\BibitemShut {NoStop}%
\bibitem [{\citenamefont {Houzet}\ and\ \citenamefont
  {Buzdin}(2007)}]{Houzet2007}%
  \BibitemOpen
  \bibfield  {author} {\bibinfo {author} {\bibfnamefont {M.}~\bibnamefont
  {Houzet}}\ and\ \bibinfo {author} {\bibfnamefont {A.~I.}\ \bibnamefont
  {Buzdin}},\ }\href {\doibase 10.1103/PhysRevB.76.060504} {\bibfield
  {journal} {\bibinfo  {journal} {Phys. Rev. B}\ }\textbf {\bibinfo {volume}
  {76}},\ \bibinfo {pages} {060504} (\bibinfo {year} {2007})}\BibitemShut
  {NoStop}%
\bibitem [{\citenamefont {Gingrich}\ \emph {et~al.}(2012)\citenamefont
  {Gingrich}, \citenamefont {Quarterman}, \citenamefont {Wang}, \citenamefont
  {Loloee}, \citenamefont {Pratt},\ and\ \citenamefont {Birge}}]{Gingrich2012}%
  \BibitemOpen
  \bibfield  {author} {\bibinfo {author} {\bibfnamefont {E.~C.}\ \bibnamefont
  {Gingrich}}, \bibinfo {author} {\bibfnamefont {P.}~\bibnamefont
  {Quarterman}}, \bibinfo {author} {\bibfnamefont {Y.}~\bibnamefont {Wang}},
  \bibinfo {author} {\bibfnamefont {R.}~\bibnamefont {Loloee}}, \bibinfo
  {author} {\bibfnamefont {W.~P.}\ \bibnamefont {Pratt}}, \ and\ \bibinfo
  {author} {\bibfnamefont {N.~O.}\ \bibnamefont {Birge}},\ }\href {\doibase
  10.1103/PhysRevB.86.224506} {\bibfield  {journal} {\bibinfo  {journal} {Phys.
  Rev. B}\ }\textbf {\bibinfo {volume} {86}},\ \bibinfo {pages} {224506}
  (\bibinfo {year} {2012})}\BibitemShut {NoStop}%
\bibitem [{\citenamefont {Martinez}\ \emph {et~al.}(2016)\citenamefont
  {Martinez}, \citenamefont {Pratt},\ and\ \citenamefont
  {Birge}}]{Martinez2016}%
  \BibitemOpen
  \bibfield  {author} {\bibinfo {author} {\bibfnamefont {W.~M.}\ \bibnamefont
  {Martinez}}, \bibinfo {author} {\bibfnamefont {W.~P.}\ \bibnamefont {Pratt}},
  \ and\ \bibinfo {author} {\bibfnamefont {N.~O.}\ \bibnamefont {Birge}},\
  }\href {\doibase 10.1103/PhysRevLett.116.077001} {\bibfield  {journal}
  {\bibinfo  {journal} {Phys. Rev. Lett.}\ }\textbf {\bibinfo {volume} {116}},\
  \bibinfo {pages} {077001} (\bibinfo {year} {2016})}\BibitemShut {NoStop}%
\bibitem [{\citenamefont {Eschrig}(2011)}]{Eschrig2011}%
  \BibitemOpen
  \bibfield  {author} {\bibinfo {author} {\bibfnamefont {M.}~\bibnamefont
  {Eschrig}},\ }\href {\doibase 10.1063/1.3541944} {\bibfield  {journal}
  {\bibinfo  {journal} {Phys. Today}\ }\textbf {\bibinfo {volume} {64}},\
  \bibinfo {pages} {43} (\bibinfo {year} {2011})}\BibitemShut {NoStop}%
\bibitem [{\citenamefont {Volkov}\ and\ \citenamefont
  {Efetov}(2010)}]{Volkov2010}%
  \BibitemOpen
  \bibfield  {author} {\bibinfo {author} {\bibfnamefont {A.~F.}\ \bibnamefont
  {Volkov}}\ and\ \bibinfo {author} {\bibfnamefont {K.~B.}\ \bibnamefont
  {Efetov}},\ }\href {\doibase 10.1103/PhysRevB.81.144522} {\bibfield
  {journal} {\bibinfo  {journal} {Phys. Rev. B}\ }\textbf {\bibinfo {volume}
  {81}},\ \bibinfo {pages} {144522} (\bibinfo {year} {2010})}\BibitemShut
  {NoStop}%
\bibitem [{\citenamefont {Trifunovic}\ and\ \citenamefont
  {Radovi\ifmmode~\acute{c}\else \'{c}\fi{}}(2010)}]{Trifunovic2010}%
  \BibitemOpen
  \bibfield  {author} {\bibinfo {author} {\bibfnamefont {L.}~\bibnamefont
  {Trifunovic}}\ and\ \bibinfo {author} {\bibfnamefont {Z.}~\bibnamefont
  {Radovi\ifmmode~\acute{c}\else \'{c}\fi{}}},\ }\href {\doibase
  10.1103/PhysRevB.82.020505} {\bibfield  {journal} {\bibinfo  {journal} {Phys.
  Rev. B}\ }\textbf {\bibinfo {volume} {82}},\ \bibinfo {pages} {020505}
  (\bibinfo {year} {2010})}\BibitemShut {NoStop}%
\bibitem [{\citenamefont {Gingrich}(2014)}]{GingrichThesis2014}%
  \BibitemOpen
  \bibfield  {author} {\bibinfo {author} {\bibfnamefont {E.~C.}\ \bibnamefont
  {Gingrich}},\ }\emph {\bibinfo {title} {Phase control of the spin-triplet
  state in \protect{S/F/S} Josephson junctions}},\ \href@noop {} {Ph.D.
  thesis},\ \bibinfo  {school} {Michigan State University} (\bibinfo {year}
  {2014})\BibitemShut {NoStop}%
\bibitem [{Sch()}]{SchneiderPrivatecomm}%
  \BibitemOpen
  \href@noop {} {}\bibinfo {note} {We thank Mike Schneider who first suggested
  we try a SAF with PMA.}\BibitemShut {Stop}%
\bibitem [{\citenamefont {Chang}\ \emph {et~al.}(2013)\citenamefont {Chang},
  \citenamefont {Canizo-Cabrera}, \citenamefont {Garcia-Vazquez}, \citenamefont
  {Chang},\ and\ \citenamefont {Wu}}]{Chang2013}%
  \BibitemOpen
  \bibfield  {author} {\bibinfo {author} {\bibfnamefont {Y.-J.}\ \bibnamefont
  {Chang}}, \bibinfo {author} {\bibfnamefont {A.}~\bibnamefont
  {Canizo-Cabrera}}, \bibinfo {author} {\bibfnamefont {V.}~\bibnamefont
  {Garcia-Vazquez}}, \bibinfo {author} {\bibfnamefont {Y.-H.}\ \bibnamefont
  {Chang}}, \ and\ \bibinfo {author} {\bibfnamefont {T.-H.}\ \bibnamefont
  {Wu}},\ }\href {\doibase 10.1063/1.4799974} {\bibfield  {journal} {\bibinfo
  {journal} {J. App. Phys.}\ }\textbf {\bibinfo {volume} {113}},\ \bibinfo
  {pages} {17B909} (\bibinfo {year} {2013})}\BibitemShut {NoStop}%
\bibitem [{\citenamefont {Lee}\ \emph {et~al.}(2016)\citenamefont {Lee},
  \citenamefont {An}, \citenamefont {Yang}, \citenamefont {Park}, \citenamefont
  {Chung},\ and\ \citenamefont {Hong}}]{Lee2016}%
  \BibitemOpen
  \bibfield  {author} {\bibinfo {author} {\bibfnamefont {J.-B.}\ \bibnamefont
  {Lee}}, \bibinfo {author} {\bibfnamefont {G.-G.}\ \bibnamefont {An}},
  \bibinfo {author} {\bibfnamefont {S.-M.}\ \bibnamefont {Yang}}, \bibinfo
  {author} {\bibfnamefont {H.-S.}\ \bibnamefont {Park}}, \bibinfo {author}
  {\bibfnamefont {W.-S.}\ \bibnamefont {Chung}}, \ and\ \bibinfo {author}
  {\bibfnamefont {J.-P.}\ \bibnamefont {Hong}},\ }\href@noop {} {\bibfield
  {journal} {\bibinfo  {journal} {Scientific Reports}\ }\textbf {\bibinfo
  {volume} {6}},\ \bibinfo {pages} {21324} (\bibinfo {year}
  {2016})}\BibitemShut {NoStop}%
\bibitem [{\citenamefont {Smith}\ \emph {et~al.}(2008)\citenamefont {Smith},
  \citenamefont {Parekh}, \citenamefont {Chunsheng}, \citenamefont {Zhang},
  \citenamefont {Donner}, \citenamefont {Lee}, \citenamefont {Khizroev},\ and\
  \citenamefont {Litvinov}}]{Smith2008}%
  \BibitemOpen
  \bibfield  {author} {\bibinfo {author} {\bibfnamefont {D.}~\bibnamefont
  {Smith}}, \bibinfo {author} {\bibfnamefont {V.}~\bibnamefont {Parekh}},
  \bibinfo {author} {\bibfnamefont {E.}~\bibnamefont {Chunsheng}}, \bibinfo
  {author} {\bibfnamefont {S.}~\bibnamefont {Zhang}}, \bibinfo {author}
  {\bibfnamefont {W.}~\bibnamefont {Donner}}, \bibinfo {author} {\bibfnamefont
  {T.~R.}\ \bibnamefont {Lee}}, \bibinfo {author} {\bibfnamefont
  {S.}~\bibnamefont {Khizroev}}, \ and\ \bibinfo {author} {\bibfnamefont
  {D.}~\bibnamefont {Litvinov}},\ }\href {\doibase 10.1063/1.2837049}
  {\bibfield  {journal} {\bibinfo  {journal} {J. App. Phys.}\ }\textbf
  {\bibinfo {volume} {103}},\ \bibinfo {pages} {023920} (\bibinfo {year}
  {2008})}\BibitemShut {NoStop}%
\bibitem [{\citenamefont {Baek}\ \emph {et~al.}(2014)\citenamefont {Baek},
  \citenamefont {Rippard}, \citenamefont {Benz}, \citenamefont {Russek},\ and\
  \citenamefont {Dresselhaus}}]{Baek2014}%
  \BibitemOpen
  \bibfield  {author} {\bibinfo {author} {\bibfnamefont {B.}~\bibnamefont
  {Baek}}, \bibinfo {author} {\bibfnamefont {W.~H.}\ \bibnamefont {Rippard}},
  \bibinfo {author} {\bibfnamefont {S.~P.}\ \bibnamefont {Benz}}, \bibinfo
  {author} {\bibfnamefont {S.~E.}\ \bibnamefont {Russek}}, \ and\ \bibinfo
  {author} {\bibfnamefont {P.~D.}\ \bibnamefont {Dresselhaus}},\ }\href
  {\doibase 10.1038/ncomms4888} {\bibfield  {journal} {\bibinfo  {journal}
  {Nature Commun.}\ }\textbf {\bibinfo {volume} {5}},\ \bibinfo {pages} {3888}
  (\bibinfo {year} {2014})}\BibitemShut {NoStop}%
\bibitem [{\citenamefont {Gingrich}\ \emph {et~al.}(2016)\citenamefont
  {Gingrich}, \citenamefont {Niedzielski}, \citenamefont {Glick}, \citenamefont
  {Wang}, \citenamefont {Miller}, \citenamefont {Loloee}, \citenamefont {{Pratt
  Jr}},\ and\ \citenamefont {Birge}}]{Gingrich2016}%
  \BibitemOpen
  \bibfield  {author} {\bibinfo {author} {\bibfnamefont {E.~C.}\ \bibnamefont
  {Gingrich}}, \bibinfo {author} {\bibfnamefont {B.~M.}\ \bibnamefont
  {Niedzielski}}, \bibinfo {author} {\bibfnamefont {J.~A.}\ \bibnamefont
  {Glick}}, \bibinfo {author} {\bibfnamefont {Y.}~\bibnamefont {Wang}},
  \bibinfo {author} {\bibfnamefont {D.~L.}\ \bibnamefont {Miller}}, \bibinfo
  {author} {\bibfnamefont {R.}~\bibnamefont {Loloee}}, \bibinfo {author}
  {\bibfnamefont {W.~P.}\ \bibnamefont {{Pratt Jr}}}, \ and\ \bibinfo {author}
  {\bibfnamefont {N.~O.}\ \bibnamefont {Birge}},\ }\href@noop {} {\bibfield
  {journal} {\bibinfo  {journal} {Nat. Phys.}\ }\textbf {\bibinfo {volume}
  {12}},\ \bibinfo {pages} {564} (\bibinfo {year} {2016})}\BibitemShut
  {NoStop}%
\bibitem [{\citenamefont {Niedzielski}\ \emph {et~al.}(2017)\citenamefont
  {Niedzielski}, \citenamefont {Bertus}, \citenamefont {Glick}, \citenamefont
  {Loloee}, \citenamefont {Pratt},\ and\ \citenamefont
  {Birge}}]{Niedzielski2017}%
  \BibitemOpen
  \bibfield  {author} {\bibinfo {author} {\bibfnamefont {B.~M.}\ \bibnamefont
  {Niedzielski}}, \bibinfo {author} {\bibfnamefont {T.~J.}\ \bibnamefont
  {Bertus}}, \bibinfo {author} {\bibfnamefont {J.~A.}\ \bibnamefont {Glick}},
  \bibinfo {author} {\bibfnamefont {R.}~\bibnamefont {Loloee}}, \bibinfo
  {author} {\bibfnamefont {W.~P.~J.}\ \bibnamefont {Pratt}}, \ and\ \bibinfo
  {author} {\bibfnamefont {N.~O.}\ \bibnamefont {Birge}},\ }\href@noop {}
  {\bibfield  {journal} {\bibinfo  {journal} {arXiv:1709.04815}\ } (\bibinfo
  {year} {2017})}\BibitemShut {NoStop}%
\bibitem [{\citenamefont {Baek}\ \emph {et~al.}(2017)\citenamefont {Baek},
  \citenamefont {Schneider}, \citenamefont {Pufall},\ and\ \citenamefont
  {Rippard}}]{Baek2017}%
  \BibitemOpen
  \bibfield  {author} {\bibinfo {author} {\bibfnamefont {B.}~\bibnamefont
  {Baek}}, \bibinfo {author} {\bibfnamefont {M.~L.}\ \bibnamefont {Schneider}},
  \bibinfo {author} {\bibfnamefont {M.~R.}\ \bibnamefont {Pufall}}, \ and\
  \bibinfo {author} {\bibfnamefont {W.~H.}\ \bibnamefont {Rippard}},\ }\href
  {\doibase 10.1103/PhysRevApplied.7.064013} {\bibfield  {journal} {\bibinfo
  {journal} {Phys. Rev. Appl.}\ }\textbf {\bibinfo {volume} {7}},\ \bibinfo
  {pages} {064013} (\bibinfo {year} {2017})}\BibitemShut {NoStop}%
\bibitem [{\citenamefont {Wang}\ \emph {et~al.}(2012)\citenamefont {Wang},
  \citenamefont {{Pratt Jr}},\ and\ \citenamefont {Birge}}]{Wang2012}%
  \BibitemOpen
  \bibfield  {author} {\bibinfo {author} {\bibfnamefont {Y.}~\bibnamefont
  {Wang}}, \bibinfo {author} {\bibfnamefont {W.~P.}\ \bibnamefont {{Pratt
  Jr}}}, \ and\ \bibinfo {author} {\bibfnamefont {N.~O.}\ \bibnamefont
  {Birge}},\ }\href {\doibase 10.1103/PhysRevB.85.214522} {\bibfield  {journal}
  {\bibinfo  {journal} {Phys. Rev. B}\ }\textbf {\bibinfo {volume} {85}},\
  \bibinfo {pages} {214522} (\bibinfo {year} {2012})}\BibitemShut {NoStop}%
\bibitem [{\citenamefont {Thomas}\ \emph {et~al.}(1998)\citenamefont {Thomas},
  \citenamefont {Ulmer},\ and\ \citenamefont {Ketterson}}]{Thomas1998}%
  \BibitemOpen
  \bibfield  {author} {\bibinfo {author} {\bibfnamefont {C.~D.}\ \bibnamefont
  {Thomas}}, \bibinfo {author} {\bibfnamefont {M.~P.}\ \bibnamefont {Ulmer}}, \
  and\ \bibinfo {author} {\bibfnamefont {J.~B.}\ \bibnamefont {Ketterson}},\
  }\href@noop {} {\bibfield  {journal} {\bibinfo  {journal} {J. App. Phys.}\
  }\textbf {\bibinfo {volume} {84}},\ \bibinfo {pages} {364} (\bibinfo {year}
  {1998})}\BibitemShut {NoStop}%
\bibitem [{\citenamefont {Kohlstedt}\ \emph {et~al.}(1996)\citenamefont
  {Kohlstedt}, \citenamefont {König}, \citenamefont {Henne}, \citenamefont
  {Thyssen},\ and\ \citenamefont {Caputo}}]{Kohlstedt1996}%
  \BibitemOpen
  \bibfield  {author} {\bibinfo {author} {\bibfnamefont {H.}~\bibnamefont
  {Kohlstedt}}, \bibinfo {author} {\bibfnamefont {F.}~\bibnamefont {König}},
  \bibinfo {author} {\bibfnamefont {P.}~\bibnamefont {Henne}}, \bibinfo
  {author} {\bibfnamefont {N.}~\bibnamefont {Thyssen}}, \ and\ \bibinfo
  {author} {\bibfnamefont {P.}~\bibnamefont {Caputo}},\ }\href@noop {}
  {\bibfield  {journal} {\bibinfo  {journal} {J. App. Phys.}\ }\textbf
  {\bibinfo {volume} {80}},\ \bibinfo {pages} {5512} (\bibinfo {year}
  {1996})}\BibitemShut {NoStop}%
\bibitem [{\citenamefont {Kotula}\ and\ \citenamefont
  {Keenan}(2006)}]{Kotula2006b}%
  \BibitemOpen
  \bibfield  {author} {\bibinfo {author} {\bibfnamefont {P.~G.}\ \bibnamefont
  {Kotula}}\ and\ \bibinfo {author} {\bibfnamefont {M.~R.}\ \bibnamefont
  {Keenan}},\ }\href@noop {} {\bibfield  {journal} {\bibinfo  {journal}
  {Micros. and Microanal.}\ }\textbf {\bibinfo {volume} {12}},\ \bibinfo
  {pages} {538} (\bibinfo {year} {2006})}\BibitemShut {NoStop}%
\bibitem [{\citenamefont {Edmunds}\ \emph {et~al.}(1980)\citenamefont
  {Edmunds}, \citenamefont {Pratt},\ and\ \citenamefont
  {Rowlands}}]{Edmunds1980}%
  \BibitemOpen
  \bibfield  {author} {\bibinfo {author} {\bibfnamefont {D.~L.}\ \bibnamefont
  {Edmunds}}, \bibinfo {author} {\bibfnamefont {W.~P.}\ \bibnamefont {Pratt}},
  \ and\ \bibinfo {author} {\bibfnamefont {J.~A.}\ \bibnamefont {Rowlands}},\
  }\href {\doibase 10.1063/1.1136116} {\bibfield  {journal} {\bibinfo
  {journal} {Rev. Sci. Instrum.}\ }\textbf {\bibinfo {volume} {51}},\ \bibinfo
  {pages} {1516} (\bibinfo {year} {1980})}\BibitemShut {NoStop}%
\bibitem [{\citenamefont {Barone}\ and\ \citenamefont
  {Patern{\`o}}(1982)}]{Barone1982}%
  \BibitemOpen
  \bibfield  {author} {\bibinfo {author} {\bibfnamefont {A.}~\bibnamefont
  {Barone}}\ and\ \bibinfo {author} {\bibfnamefont {G.}~\bibnamefont
  {Patern{\`o}}},\ }\href@noop {} {\emph {\bibinfo {title} {{Physics and
  applications of the Josephson effect}}}}\ (\bibinfo  {publisher} {Wiley},\
  \bibinfo {year} {1982})\BibitemShut {NoStop}%
\bibitem [{\citenamefont {Ambegaokar}\ and\ \citenamefont
  {Halperin}(1969)}]{AmbegaokarHalperin1969}%
  \BibitemOpen
  \bibfield  {author} {\bibinfo {author} {\bibfnamefont {V.}~\bibnamefont
  {Ambegaokar}}\ and\ \bibinfo {author} {\bibfnamefont {B.~I.}\ \bibnamefont
  {Halperin}},\ }\href@noop {} {\bibfield  {journal} {\bibinfo  {journal}
  {Phys. Rev. Lett.}\ }\textbf {\bibinfo {volume} {22}},\ \bibinfo {pages}
  {1364} (\bibinfo {year} {1969})}\BibitemShut {NoStop}%
\bibitem [{\citenamefont {Ivanchenko}\ and\ \citenamefont
  {Zil'berman}(1968)}]{IvanchenkoZilberman1969}%
  \BibitemOpen
  \bibfield  {author} {\bibinfo {author} {\bibfnamefont {Y.~M.}\ \bibnamefont
  {Ivanchenko}}\ and\ \bibinfo {author} {\bibfnamefont {L.~A.}\ \bibnamefont
  {Zil'berman}},\ }\href@noop {} {\bibfield  {journal} {\bibinfo  {journal}
  {Zh. Eksp. Teor. Fiz}\ }\textbf {\bibinfo {volume} {55}},\ \bibinfo {pages}
  {2395} (\bibinfo {year} {1968})},\ \bibinfo {note} {[\textit{Soviet Physics
  J. Exp. Theor. Phys.}, \textbf{28}, 6, (1969)]}\BibitemShut {NoStop}%
\bibitem [{\citenamefont {Baek}()}]{Baek_privatecommunication}%
  \BibitemOpen
  \bibfield  {author} {\bibinfo {author} {\bibfnamefont {B.}~\bibnamefont
  {Baek}},\ }\href@noop {} {}\bibinfo {note} {Private communication. One may
  need 400 mT to initialize the magnetization direction of a thin Ni
  nanomagnet.}\BibitemShut {Stop}%
\bibitem [{\citenamefont {Ryazanov}(1999)}]{Ryazanov1999}%
  \BibitemOpen
  \bibfield  {author} {\bibinfo {author} {\bibfnamefont {V.~V.}\ \bibnamefont
  {Ryazanov}},\ }\href {http://stacks.iop.org/1063-7869/42/i=8/a=A06}
  {\bibfield  {journal} {\bibinfo  {journal} {Physics-Uspekhi}\ }\textbf
  {\bibinfo {volume} {42}},\ \bibinfo {pages} {825} (\bibinfo {year}
  {1999})}\BibitemShut {NoStop}%
\bibitem [{\citenamefont {Korucu}(2016)}]{Korucu-unpublished}%
  \BibitemOpen
  \bibfield  {author} {\bibinfo {author} {\bibfnamefont {D.}~\bibnamefont
  {Korucu}},\ }\href@noop {} {\bibfield  {journal} {\bibinfo  {journal}
  {Unpublished}\ } (\bibinfo {year} {2016})}\BibitemShut {NoStop}%
\bibitem [{\citenamefont {Nguyen}\ \emph {et~al.}(2010)\citenamefont {Nguyen},
  \citenamefont {Acharyya}, \citenamefont {Huey}, \citenamefont {Richard},
  \citenamefont {Loloee}, \citenamefont {Pratt}, \citenamefont {Bass},
  \citenamefont {Wang},\ and\ \citenamefont {Xia}}]{Nguyen2010}%
  \BibitemOpen
  \bibfield  {author} {\bibinfo {author} {\bibfnamefont {H.~Y.~T.}\
  \bibnamefont {Nguyen}}, \bibinfo {author} {\bibfnamefont {R.}~\bibnamefont
  {Acharyya}}, \bibinfo {author} {\bibfnamefont {E.}~\bibnamefont {Huey}},
  \bibinfo {author} {\bibfnamefont {B.}~\bibnamefont {Richard}}, \bibinfo
  {author} {\bibfnamefont {R.}~\bibnamefont {Loloee}}, \bibinfo {author}
  {\bibfnamefont {W.~P.}\ \bibnamefont {Pratt}}, \bibinfo {author}
  {\bibfnamefont {J.}~\bibnamefont {Bass}}, \bibinfo {author} {\bibfnamefont
  {S.}~\bibnamefont {Wang}}, \ and\ \bibinfo {author} {\bibfnamefont
  {K.}~\bibnamefont {Xia}},\ }\href {\doibase 10.1103/PhysRevB.82.220401}
  {\bibfield  {journal} {\bibinfo  {journal} {Phys. Rev. B}\ }\textbf {\bibinfo
  {volume} {82}},\ \bibinfo {pages} {220401} (\bibinfo {year}
  {2010})}\BibitemShut {NoStop}%
\bibitem [{\citenamefont {Pollard}\ \emph {et~al.}(2017)\citenamefont
  {Pollard}, \citenamefont {Garlow}, \citenamefont {Yu}, \citenamefont {Wang},
  \citenamefont {Zhu},\ and\ \citenamefont {Yang}}]{Pollard2017}%
  \BibitemOpen
  \bibfield  {author} {\bibinfo {author} {\bibfnamefont {S.~D.}\ \bibnamefont
  {Pollard}}, \bibinfo {author} {\bibfnamefont {J.~A.}\ \bibnamefont {Garlow}},
  \bibinfo {author} {\bibfnamefont {J.}~\bibnamefont {Yu}}, \bibinfo {author}
  {\bibfnamefont {Z.}~\bibnamefont {Wang}}, \bibinfo {author} {\bibfnamefont
  {Y.}~\bibnamefont {Zhu}}, \ and\ \bibinfo {author} {\bibfnamefont
  {H.}~\bibnamefont {Yang}},\ }\href {http://dx.doi.org/10.1038/ncomms14761}
  {\bibfield  {journal} {\bibinfo  {journal} {Nat. Comm.}\ }\textbf {\bibinfo
  {volume} {8}},\ \bibinfo {pages} {14761} (\bibinfo {year}
  {2017})}\BibitemShut {NoStop}%
\end{thebibliography}%
\end{document}